\documentclass[fleqn,usenatbib]{mnras}

\usepackage{newtxtext,newtxmath}

\usepackage[T1]{fontenc}
\usepackage{caption}
\usepackage{subcaption}
\DeclareRobustCommand{\VAN}[3]{#2}
\let\VANthebibliography\thebibliography
\def\thebibliography{\DeclareRobustCommand{\VAN}[3]{##3}\VANthebibliography}

\usepackage{comment}
\usepackage{graphicx}	
\usepackage{amsmath}	

\newcommand\swift{{\it Swift}\,}
\newcommand\fermi{{\it Fermi}\,}

\title[GRB211211A: Precursor periodicity]{Prompt Periodicity in the GRB 211211A Precursor: Black-hole or magnetar engine?}

\author[G. P. Lamb et al.]{Gavin P. Lamb\href{https://orcid.org/0000-0001-5169-4143}{\includegraphics[scale=0.5]{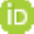}}$^{1}$\thanks{E-mail: g.p.lamb@ljmu.ac.uk},
Thomas Baxter\href{https://orcid.org/0009-0004-4738-9022}{\includegraphics[scale=0.5]{ORCIDiD_icon16x16-eps-converted-to.pdf}}$^{1}$,
Conor M. B. Omand\href{https://orcid.org/0000-0002-9646-8710}{\includegraphics[scale=0.5]{ORCIDiD_icon16x16-eps-converted-to.pdf}}$^{1}$,
Dimple\href{https://orcid.org/0000-0001-9868-9042}{\includegraphics[scale=0.5]{ORCIDiD_icon16x16-eps-converted-to.pdf}}$^{2}$,
Zoë McGrath\href{https://orcid.org/0009-0006-0726-1328}{\includegraphics[scale=0.5]{ORCIDiD_icon16x16-eps-converted-to.pdf}}$^{1}$,
\newauthor 
Cairns Turnbull\href{https://orcid.org/0009-0002-7044-5358}{\includegraphics[scale=0.5]{ORCIDiD_icon16x16-eps-converted-to.pdf}}$^{1}$,
Eric Burns\href{https://orcid.org/0000-0002-2942-3379}{\includegraphics[scale=0.5]{ORCIDiD_icon16x16-eps-converted-to.pdf}}$^{3}$,
Hamid Hamidani\href{https://orcid.org/0000-0003-2866-4522}{\includegraphics[scale=0.5]{ORCIDiD_icon16x16-eps-converted-to.pdf}}$^{4}$,
Ilya Mandel\href{https://orcid.org/0000-0002-6134-8946}{\includegraphics[scale=0.5]{ORCIDiD_icon16x16-eps-converted-to.pdf}}$^{5,6}$,
Kim L. Page\href{https://orcid.org/0000-0001-5624-2613}{\includegraphics[scale=0.5]{ORCIDiD_icon16x16-eps-converted-to.pdf}}$^{7}$,
\newauthor 
Stephan Rosswog\href{https://orcid.org/0000-0002-3833-8520}{\includegraphics[scale=0.5]{ORCIDiD_icon16x16-eps-converted-to.pdf}}$^{8,9}$,
Nikhil Sarin\href{https://orcid.org/0000-0003-2700-1030}{\includegraphics[scale=0.5]{ORCIDiD_icon16x16-eps-converted-to.pdf}}$^{9,10}$,
Andrew Blain\href{https://orcid.org/0000-0001-7489-5167}{\includegraphics[scale=0.5]{ORCIDiD_icon16x16-eps-converted-to.pdf}}$^{6}$,
Laurence Datrier\href{https://orcid.org/0000-0002-0290-3129}{\includegraphics[scale=0.5]{ORCIDiD_icon16x16-eps-converted-to.pdf}}$^{11}$,
Shiho Kobayashi\href{https://orcid.org/0000-0001-8172-4411}{\includegraphics[scale=0.5]{ORCIDiD_icon16x16-eps-converted-to.pdf}}$^{1}$,
\newauthor 
Andrew Levan\href{https://orcid.org/0000-0001-7821-9369}{\includegraphics[scale=0.5]{ORCIDiD_icon16x16-eps-converted-to.pdf}}$^{12}$,
Rhaana Starling\href{https://orcid.org/0000-0001-5803-2038}{\includegraphics[scale=0.5]{ORCIDiD_icon16x16-eps-converted-to.pdf}}$^{7}$,
Benjamin Gompertz\href{https://orcid.org/0000-0002-5826-0548}{\includegraphics[scale=0.5]{ORCIDiD_icon16x16-eps-converted-to.pdf}}$^{2}$,
Nusrin Habeeb$^{7}$,
Khang Nguyen$^{1}$,
\newauthor
Nial Tanvir\href{https://orcid.org/0000-0003-3274-6336}{\includegraphics[scale=0.5]{ORCIDiD_icon16x16-eps-converted-to.pdf}}$^{7}$.
\\
$^{1}$Astrophysics Research Institute, Liverpool John Moores University, IC2 Liverpool Science Park, 146 Brownlow Hill, Liverpool, L3 5RF, UK\\
$^{2}$School of Physics and Astronomy, University of Birmingham, Edgbaston, Birmingham, B15 2TT, UK\\
$^{3}$Department of Physics and Astronomy, Louisiana State University, Baton Rouge, Louisiana 70803, USA\\
$^{4}$Astronomical Institute, Graduate School of Science, Tohoku University, Sendai 980-8578, Japan\\
$^5$School of Physics and Astronomy, Monash University, Clayton, Victoria 3800, Australia\\
$^6$ARC Centre of Excellence for Gravitational Wave Discovery -- OzGrav, Australia\\
$^{7}$School of Physics and Astronomy, University of Leicester, University Road, Leicester, LE1 7RH \\
$^{8}$University of Hamburg, Hamburger Sternwarte, Gojenbergsweg 112, 21029, Hamburg, Germany\\
$^{9}$The Oskar Klein Centre, Department of Physics, Stockholm University, AlbaNova, SE-106 91 Stockholm, Sweden\\
$^{10}$Nordita,  Stockholm University and KTH Royal Institute of Technology
Hannes Alfvéns väg 12, SE-106 91 Stockholm, Sweden\\
$^{11}$California State University - Fullerton, 800 N State College Blvd, Fullerton CA 92831, USA\\
$^{12}$Radboud University Nijmegen: Nijmegen, Gelderland, NL\\
}

\date{Accepted XXX. Received YYY; in original form `Month' 2025}

\pubyear{\the\year{}}

\begin{document}
\label{firstpage}
\pagerange{\pageref{firstpage}--\pageref{lastpage}}
\maketitle

\begin{abstract}
The merger origin long GRB\,211211A was a class (re-)defining event. A precursor was identified with a $\sim 1$\,s separation from the main burst, as well as a claimed candidate quasi-periodic oscillation (QPO) with a frequency $\sim20$\,Hz. Here, we explore the implications of the precursor, assuming the quasi-periodicity is real. The precursor variability timescale requires relativistic motion with a Lorentz factor $\Gamma\gtrsim80$, and implies an engine driven jetted outflow. The declining amplitude of the consecutive pulses requires an episodic engine with an `on/off' cycle consistent with the QPO. For a black-hole central engine, the QPO can have its origin in Lense-Thirring precession of the inner disk at $\sim6-9$\,$r_g$ (gravitational radii) for a mass $M_\bullet\leq4.5$\,M$_{\sun}$, and $\lesssim 7$\,$r_g$ for $M_\bullet>4.5$\,M$_{\sun}$ and dimensionless spin $\chi\sim 0.3 - 0.9$. Alternatively, at a disk density of $\sim10^{8 - 12}$\,g\,cm$^{-3}$, the required magnetic field strength for a QPO via magnetohydrodynamic effects will be on the order $B\sim10^{12 - 14}$\,G. If the central engine is a short lived magnetar or hypermassive neutron star, then a low-frequency QPO can be produced via instabilities within the disk at a radius of $\sim20 - 70$\,km, for a disk density $\sim10^{9 - 12}$\,g\,cm$^{-3}$ and magnetic field $\gtrsim10^{13 - 14}$\,G. The QPO cannot be coupled to the neutron star spin, as the co-rotation radius is beyond the scale of the disk. Neither engine can be ruled out -- 
however, we favour an origin for the precursor candidate QPO as early jet-disk coupling for a neutron star -- black hole merger remnant with mass $M_\bullet>4.5$\,M$_{\sun}$. 
\end{abstract}




\section{Introduction}
Gamma-ray bursts (GRBs) are characteristically high-energy events with very short variability timescales.
The combination of high luminosity and short variability requires these sources to have ultra-relativistic velocities $\Gamma\gg10$, with Lorentz factors typically on the order of $\Gamma\sim100$ considered.

The prompt GRB emission light curves can be separated into one or more pulses \citep[e.g.,][]{kobayashi1997, norris1996, norris2005}.
Beyond these short timescale variations, there are three main phases identified within the GRB prompt emission \citep[e.g.,][]{hu2014}: a precursor \citep{koshut1995}; the main burst complex; and extended, plateau and late-flaring emission \citep{burrows2005}. 
While the main phase is usually the most energetic and well-studied, precursors (short bursts of radiation occurring before the main event) have been a compelling component of GRB prompt emission \citep{koshut1995, lazzati2005, burlon2008, burlon2009, hu2014, zhong2019, coppin2020, li2021, li2022}, with many competing explanations.
These include:
pair cascades producing a quasi-thermal X-ray pulse \citep{meszaros2001};
shock or cocoon breakout from the stellar envelope or merger ejecta \citep{ramirez-ruiz2002, nakar2017};
`spinar-like object' due to a two phase core-collapse \citep{lipunova2009};
a central engine origin that is the same as the main burst complex, for short GRB precursors \citep{troja2010};
distinct fireballs with their own afterglow \citep{nappo2014};
crustal resonances and shattering in pre-merger neutron stars \citep{tsang2012, suvorov2020, neill2022};
g-mode oscillations in pre-merger neutron stars \citep{kuon2021};
changes in the energy content of the early jet, baryonic to Poynting flux dominated \citep{wang2020};
etc.

The main burst complex is used to classify GRBs via their observed duration; short ($<2$\,s) and long ($> 2$\,s) \citep{kouveliotou1993}; alternatively, schemes put GRBs into two categories -- Type I and Type II \citep{zhang2009}.
 For GRB precursors, there is a distinction between those from Type I (typically short duration GRBs) and those from Type II (typically long-duration GRBs).
The latter are well-fit with a thermal spectrum and exhibit a smoother light curve \citep[e.g.][]{wang2007,gutierrez2024}, whereas the former are typically consistent with the emission qualities of the main burst complex \citep[e.g.,][]{troja2010}.

Amongst the GRB population are several events that challenge the established classification schemes; the most prominent and convincing of these is GRB\,211211A \citep{gompertz2023}.
Despite its long duration ($51$\,s or $34.3$\,s for \fermi-Gamma Burst Monitor (GBM) and \swift-Burst Alert Telescope (BAT) respectively), the burst was accompanied by a candidate kilonova, seen within the afterglow \citep{rastinejad2022}.
A kilonova is a thermal transient produced via neutron star mergers and is the observational smoking gun for a merger origin of GRBs \citep[see][for a review of kilonova]{metzger2020}.
At the time of discovery, this kilonova was the best-sampled example of such a transient within the afterglow of a classical GRB -- the kilonova following the neutron star merger GW170817 being the best sampled kilonova case, however, GRB\,170817A was far from typical/classical in terms of the GRB population \citep[see][for the discovery paper]{abbott2017}.

The afterglow, candidate kilonova, candidate host galaxy, and the prompt emission were thoroughly investigated \citep[e.g.,][]{waxman2022, gao2022, zhu2022, zhang2022, yang2022, mei2022, rastinejad2022, chang2023, gompertz2023, hamidani2024}.
Compounding the uniqueness of GRB\,211211A, the burst had a short ($\sim$0.2\,s) precursor with an isotropic equivalent energy, $E_{\rm pre,iso} \sim7\times10^{48}$\,erg, which exhibited apparent periodicity, with claims for a quasi-periodic oscillation (QPO) at $\sim22$\,Hz \citep{xiao2024}.
A further QPO candidate within the main burst complex, again lasting $\sim0.2$\,s, was claimed by \citet{chirenti2024}, however, their analysis found the precursor QPO claimed by \citet{xiao2024} to be below their levels of significance.
We note that \citet{xiao2024} performed multiple significance tests for the QPO, and found it to be significant by all measures.

Several progenitor models have been proposed to explain the origin of GRB\,211211A:
a neutron star-white dwarf merger with a rapidly spinning magnetar remnant \citep{yang2022,yi-zhong2024}; resonances within neutron star oceans \citep{sullivan2024}; a neutron star-black hole or binary neutron star merger \citep{barnes2023,kunert2023,zhu2022,Dimple2023}.
Notably, QPOs have been suggested as an indicator of neutron star-black hole merger remnants \citep{stone2013, li2023}.

The candidate QPO in the precursor of GRB\,211211A presents a unique opportunity to probe the mechanisms driving the early phases of GRB emission related to accretion and jet launching. 
QPOs have been observed in various astrophysical systems -- however, their presence in GRBs is rare and provides hints to the nature of the central engine (see \citet{chirenti2023} for the discovery of two QPOs within legacy catalogues of short-duration bursts GRB 910711 and 931101B, and \citet{yang2025} for further high-frequency discoveries within short GRBs). 
For the kilo-Hertz frequency QPOs in GRBs 910711 and 931101B, the origin aligns well with the expectation from neutron star mergers, and \citet{guedes2024} have proposed using this to constrain the burst redshift and infer the mass-radius relation of the neutron stars.
Our study reinvestigates the precursor and candidate, low-frequency QPO of GRB\,211211A, with an aim to characterize its properties and to explore its physical origin. 

For binary neutron star mergers, several mechanisms exist that can produce precursor-like emission before the merger \citep{fernandez2016}.
These include magnetospheric interactions between the two neutron stars -- however, these are predicted to produce flares with a maximum luminosity of $L\leq10^{44}$\,erg\,s$^{-1}$ \citep{wang2018}, several orders of magnitude fainter than the GRB\,211211A precursor.
An alternative to magnetospheric interactions are resonant shattering flares \citep{tsang2012}, where the tidal interactions between the pre-merger neutron stars can produce a resonance sufficient to shatter the crust, leading to a series of ejected shells that can collide to produce internal shocks similar to those of the GRB prompt emission but on the order of $L\leq 10^{48}$\,erg\,s$^{-1}$ \citep{neill2022}. However, the candidate QPO frequency and the temporal structure of the GRB\,211211A precursor are inconsistent with such resonant shattering flares \citep[see the addendum within][]{neill2022}.
A second long-duration merger origin burst, GRB\,230307A \citep{levan2024}, also contained a potential precursor. The resonant shattering flare scenario was invoked to explain this emission, despite the candidate precursor's relatively high luminosity (for a precursor) -- by invoking high magnetic field strengths, \citet{dichiara2023} were able to find a viable parameter space in this case.
The high magnetic field strengths would require one of the merging components to be a magnetar. While magnetar lifetime is typically considered to be shorter than the neutron star merger timescale, there is some evidence of an old population of magnetars \citep{beniamini2023}.
A final precursor model for GRB\,211211A invokes the merger of a black-hole and a massive neutron star, producing tidal resonances that can produce the observed precursor yet struggles to explain the main burst complex or kilonova \citep{sullivan2024}.
For GRB\,211211A, the various pre-merger flare models can be ruled out due to the luminosity or the temporal structure of the precursor.

We do not attempt to confirm the significance of the candidate QPO within the precursor of GRB\,211211A, but rather explore viable mechanisms for such a QPO given a compact merger origin for the GRB.
Robust tests for the significance of a QPO could involve Gaussian process modelling \citep{hubner2022} -- such a test is beyond the scope of the current work, and has some overlap with the methods employed by \citet{xiao2024} in their discovery paper.

We first discuss potential origins for a QPO in a post-merger system in \S\ref{s:QPO}.
The data used in our search for the candidate QPO are described in \S\ref{s:data}, and the results of the data analysis given in \S\ref{s:res}.
The results are discussed in the context of an accreting compact object and the various mechanisms explored in \S\ref{s:disc}, and our conclusions are made in \S\ref{s:conc}.

\section{Potential QPO origins}\label{s:QPO}
For GRBs, accretion onto a newly formed stellar mass black hole, or a magnetar (magnetized neutron star), is the most likely engine for jetted emission \citep[see e.g.,][]{fryer2019}. 
For these scenarios, the exact nature of the progenitor system is not exclusive. 
For instance, a magnetar system for GRB 211211A could result from a neutron-star merger or a white-dwarf neutron star merger \citep[e.g.,][]{yang2022}.
Similarly, a black-hole engine can be the product of either a neutron star binary or a neutron star black-hole merger.

More exotic but potentially viable models would also fall under one of these two categories. 
For example, consider white dwarf-black hole mergers; although unlikely to produce a GRB \citep{narayan2001}, if they did, they would follow a black-hole accretion scenario. 
In another case, quark stars, if produced from a neutron star merger, would follow either a black-hole or magnetar scenario, depending on the specific properties of the quark stars \citep{cheng1996}.

Meanwhile, even more exotic progenitor systems can be safely ruled out, as the origin of the prompt emission, due to the phenomenology and properties of GRB 211211A. 
Soft gamma repeaters and magnetar giant flares, for instance, are extremely short in duration and have a local Universe origin. 
Primordial black hole evaporation is expected to result in a thermal, short-duration GRB with an origin within a few parsecs \citep{cline1992}. 
Finally, nuclear goblins (probably) do not exist \citep{zwicky1974}.

Here we outline possible mechanisms for generating QPOs with such engines.

\subsection{Black-hole accretion}
For black-hole accretion disk systems,  there are several potential origins for QPOs in the emission. 
Oscillation within the precursor implies that the engine exhibits periodicity at very early times.

\subsubsection{Lense-Thirring precession}

Lense-Thirring (LT) precession is the result of frame dragging within the disk of a black-hole with spin $\chi > 0$.
The frequency of the LT precession is given by \citep{bardeen1975, kumar1985}:
\begin{equation}
    \nu_{\rm LT} = \frac{\omega_{\rm LT}}{2{\pi}} = \frac{\chi G^2M_\bullet^2}{{\pi} c^3 r_{\rm id}^3},
    \label{eq:LTfreq}
\end{equation}
where $\chi$ is the dimensionless spin of the black-hole, $M_\bullet$ is the black-hole mass, $c$ is the speed of light, $G$ is the gravitational constant, and $r_{\rm id}$ is the inner radius of the accretion disc.
In the following we will always assume that the disk mass is negligible with respect to the central mass, which should be a reasonable approximation.

\subsubsection{Diskoseismology: p- and g-modes}
Pressure and gravity modes of buoyancy waves trapped within a black-hole accretion disk can result in a QPO \citep{kato1980, wagoner1999}.
The inertial acoustic, or pressure modes (p-modes), where pressure is the restoring force, have oscillation frequencies of,
\begin{equation}
    \nu_{\rm p-mode} \simeq \frac{1}{2\pi} \sqrt{\frac{GM_\bullet}{r^3}}
    = \frac{\omega_{\rm K}(r)}{2 \pi} \equiv \nu_{\rm K}(r),
    \label{eq:p-mode}
\end{equation}
where we have denoted the Kepler angular frequency as:
\begin{equation}
\omega_{\rm K}(r)= \sqrt{\frac{GM_\bullet}{r^3}}.
\end{equation}

And gravity modes (g-modes), where gravity is the restoring force, have frequencies
\begin{equation}
    \nu_{\rm g-mode} \simeq \nu_{\rm p-mode} \frac{N_{BV}}{\kappa},
    \label{eq:g-mode}
\end{equation}
where $N_{BV}$ is the buoyancy (Brunt-V$\ddot{\rm a}$is$\ddot{\rm a}$l$\ddot{\rm a}$) frequency, and ${\kappa}$ is the epicyclic  (a ``Spirograph''-like motion) frequency.

\subsubsection{Orbital resonance}
When a QPO is related to the orbital, radial, and vertical epicyclic frequencies, these are called orbital resonances \citep{torok2005}.
These characteristic frequencies are:
\begin{equation}
    \nu_{\phi} = \nu_{\rm K}(r) .
    \label{eq:kepler}
\end{equation}
The orbital frequency has the same order as the p-mode of diskoseismology -- however, the mechanisms are distinct.
Whereas the p-mode is a local oscillation within the disk driven by pressure forces, the orbital resonance is a global phenomenon that couples to fundamental frequencies.
This non-linear coupling of fundamental frequencies can produce pairs of QPOs.

The radial and vertical frequencies are then \citep{kato2001}:
\begin{align}
    \nu_{r} &= \nu_{\phi}\sqrt{1 - \frac{6GM_\bullet}{r c^2}+\frac{8\chi GM_\bullet^{3/2}}{c^3r^{3/2}}-\frac{3\chi^2GM_\bullet^2}{c^4r^2}},\\
    \nu_{\theta} &= \nu_{\phi}\sqrt{1 - \frac{4\chi GM_\bullet}{r^{3/2} c^2}}.
    \label{eq:freqepi}
\end{align}
Where the epicyclic frequency and orbital frequency have the ratio 3/2, the system is in the 3:2 resonance, and at low frequencies, the QPO can be due to the precession frequency, $\nu_{\rm low} = \nu_\phi - \nu_{r}$.

\subsubsection{Blobs}
If the disk contains a blob, or hot spot, at a particular radius,
there will be an oscillation at the Keplerian frequency, $\nu_{\rm blob} \equiv \nu_{\phi}$, equation\,\ref{eq:kepler}.

\subsubsection{Magnetohydrodynamics process}
Magnetohydrodynamic (MHD) turbulence such as magnetorotational instabilities (MRI) can drive QPOs and episodic accretion \citep{masada2007, masada2009}.
The QPOs will have a frequency given by:
\begin{equation}
    \nu_{\rm MRI} \simeq  \frac{k_\parallel~v_A}{2\pi},
    \label{eq:MRI}
\end{equation}
where $k_\parallel$ is the wavenumber and $v_A$ is the Alfv\'en speed \citep{schekochinhin2009}.

\subsubsection{Accretion flow shock oscillations}
Shocks within the accretion flow can form in the inner disk regions.
These shocks will oscillate on the dynamical timescale such that:
\begin{equation}
    \nu_{\rm shock} \simeq \nu_{\rm K}(r_{\rm shock}) , 
\end{equation}
where $r_{\rm shock}$ is the shock radius.
As the oscillations depend on the shock location, the frequency of the oscillation can vary from a few to several hundred Hz \citep{debnath2024}.

\subsubsection{Timescale and damping limits for a black hole - disk QPO}
The general precursor temporal structure follows a fast rise with exponential decline (FRED) \citep{xiao2024}.
This FRED structure is usually explained with the rise due to shock emission, and the decline being higher latitude emission.
However, here the precursor has oscillations and pulses within the decline.

For LT precession, the short duration of the effect puts limits on the scale of the precessing disk element.
By equating the QPO duration with the system coherence time, the size of the disk region that participates in the LT precession can be found by considering the differential precession:
\begin{equation}
    \Delta r \simeq \frac{r}{3 \nu_{\rm LT} t_{\rm coh}}, 
    \label{eq:dif_tcoh}
\end{equation}
where $\Delta r$ is the size of the disk ring responsible for LT precession and $t_{\rm coh}$ is the coherence time.

Mechanisms that can mediate the damping of oscillations include \citep[see e.g.,][and references therein]{armitage2022}:
viscous damping;
disk warping;
radiative cooling;
magnetic damping;
and jet interaction.

Viscous stressing within the precessing disk region will dissipate energy, the timescale is \citep{king2007},
\begin{equation}
    t_{\rm visc} \simeq 
    \begin{cases}
        \frac{r^2}{\alpha c_s h}\hspace{0.55cm}{\rm thin~disk},\\
        \frac{r}{\alpha c_s}\hspace{0.6cm}{\rm thick~disk},       
    \end{cases} 
    \label{eq:tvisc}
\end{equation}
where $h$ is the disk scale height, 
$\alpha$ is the dimensionless viscosity parameter and $c_s \leq c/\sqrt{3}$ is the sound speed.
For the thick disk case, we have assumed $h \simeq r$. 

LT precession of the inner disk can additionally result in a warped or torn disk \citep{nealon2015}. The timescale for damping of an oscillation will be the dynamical time, $t_{\rm dyn} \propto r^{3/2} M_\bullet^{-1/2}$.
This is equivalent to the instability growth timescale, and the resultant warping or tearing will dissipate energy or break up the disk, leading to damping of the QPO amplitude.

For a precession due to a `blob' which is hot, radiative cooling will reduce the temperature and pressure, resulting in energy dissipation.
The timescale for such radiative cooling is typically equivalent to the dynamical timescale, $t_{\rm cool} \equiv t_{\rm dyn}$.
The dynamical timescale will be small for our system -- however, if the disk is optically thick at early times, then the cooling timescale will be $t_{\rm cool} > t_{\rm dyn}$.

Interaction between magnetic fields within the disk and the precessing structure can result in magnetic braking or reconnection \citep{tabone2022}.
Either would dissipate energy, with the relevant timescale being the Alfv\'en crossing time, $t_{\rm A\times} \sim h/v_A$.
However, this timescale is sensitive to the scale of the magnetic structure.

Where the precessing structure is coupled to the jet, this jet-disk interaction will extract energy.
As energy is removed from the disk, the precession will be dampened.
The timescale for this will be comparable to the observed duration, and requires the total jet power to be comparable, or greater than, the power in the disk precession.

The apparent damping of the QPO could be the result of any, or some combination, of these various mechanisms.

\subsection{Magnetar engine}
Rapidly rotating, newly formed magnetars are commonly invoked as the power source for many high energy astrophysical transients, particularly where the energy source is ambiguous.
For GRBs, short periodic signatures are potential evidence for magnetars as the engine that drives the GRB-producing jet.

\subsubsection{Alfv\'en waves within a magnetosphere}
Magnetars have multiple ways to produce QPOs.
The initial analysis of the QPO by \cite{xiao2024} invoked a magnetar process -- either pre-merger magnetosphere interactions or crustal oscillations.
The most relevant QPO production mechanism for our scenario, where a relatively low oscillation frequency is observed, on the order $\sim1$ to $100$ Hz, is via Alfv\'en waves within the magnetosphere\footnote{This process is analogous to the MHD oscillations described for black-hole accretion disks with equation\,\ref{eq:MRI}}:
\begin{equation}
    \nu_{\rm QPO}\simeq \frac{B}{r_A\sqrt{4\pi\rho}},
    \label{eq:alfenQPO}
\end{equation}
where $r_A$ is the Alfv\'en wavelength, $B$ is the magnetic field strength, and $\rho$ is the magnetospheric plasma density.
For our scenario, the natural assumption is to make the spin period of the magnetar directly related to the QPO frequency as $\nu_{\rm QPO}\sim 1/P$, where $P$ is the period. 

\subsubsection{Damping timescale of Alfv\'en waves}
Alfv\'en waves will lose energy via transfer to other modes due to nonlinear wave-wave interactions.
Such nonlinear damping is significant in highly turbulent magnetospheres, as expected for a newly formed magnetar following a neutron star merger.
The damping timescale can be approximated as:
\begin{equation}
    \tau_{\rm nl} \simeq \frac{r_A}{v_A \delta B/B},
    \label{eq:damp}
\end{equation}
where $v_A = B/\sqrt{4\pi \rho}$ is the Alfv\'en speed, and $0<\delta B/B \leq 1$ is the fractional amplitude of the Alfv\'en wave (i.e., the ratio of the Alfv\'en wave amplitude to the background magnetic field).
The damping timescale for a magnetar with magnetic field $B\sim10^{14}$\,G, plasma density on the order $\rho \sim 10^{12}$\,g\,cm$^{-3}$, and $\delta B/B\sim 0.22$, will have a timescale of $\sim0.2$\,s, on the same order as the precursor.
These parameter values are order-of-magnitude estimates given the expectation for these scenarios and demonstrate that the damping timescale for such a QPO system is consistent with the observed duration.

\section{Data}\label{s:data}
We reanalysed the time series data for the sub-second precursor within the initial burst complex of GRB\,211211A. 
Our reanalysis aims to identify\footnote{Rigorous significance testing for this $22.5$\,Hz QPO was performed via multiple methods by \cite{xiao2024}, we do not present the repetition of that analysis here.} the quasi-periodic signal identified by \cite{xiao2024} in the $\sim0.2$\,s precursor of the long-duration GRB\,211211A.
We used publicly available $15 - 350$\,keV data of the GRB event from two telescopes in our study.
GRB\,211211A has a redshift $z=0.076$ \citep{rastinejad2022}.

\begin{table}
    \centering
    \begin{tabular}{c|c|c}
        label & energy range (keV) & see figure \\
        \hline
        band 1 & 15-25 & [\ref{fig:FERMI},\ref{fig:BAT},\ref{fig:LSHS_low}]\\
        band 2 & 25-50 & [\ref{fig:FERMI},\ref{fig:BAT}]\\
        band 3 & 50-100 & [\ref{fig:FERMI},\ref{fig:BAT}]\\
        band 4 & 100-350 & [\ref{fig:FERMI},\ref{fig:BAT}]\\
        low & 15-25 & [\ref{fig:LSHS_low}]\\
        high & 25-350 & [\ref{fig:LSHS_low},\ref{fig:pulse_fit}]\\
        all & 15-350 & [\ref{fig:FERMI_v_BAT},\ref{fig:APP_A1},\ref{fig:welch}]\\
    \end{tabular}
    \caption{Reference labels used throughout for the seven light curve options. The spectral energy range for each light curve and the figures where this data was used.}
    \label{tab:LCs}
\end{table}

\subsection{Swift-BAT}
We used the standard {\sc HEASoft} {\sc batgrbproduct} and {\sc batbinevt} tools to extract light curves for the \swift-BAT prompt emission over the default energy bands of $15-25$, $25-50$, $50-100$, and $100-350$\,keV. The data were re-binned to 4ms bins for convenient short timescale/higher frequency sampling and to accurately resolve the precursor. We summed all energy channels to create a `total' energy light curve, in addition to low- and high-energy light curves with ranges $15 - 25$ and $25 - 350$\,keV, as well as the default bands, giving seven light curve options (see Table\,\ref{tab:LCs}).

\subsection{Fermi-GBM}
We used the time-tagged event (TTE) data\footnote{\href{https://fermi.gsfc.nasa.gov/ssc/data/access/gbm/}{\fermi-GBM data}.} for n2 and na, the highest signal-to-noise \fermi-GBM detectors for GRB\,211211A.
The TTE data have 128 energy channels with $\sim2~\mu$s precision, and we binned this using the {\sc GDT-core} tools to 4\,ms bins.
We further divided the data as for \swift-BAT into distinct bands listed in Table\,\ref{tab:LCs}.

\section{Analysis and Results}\label{s:res}
The data analysis methods for our QPO search, pulse identification, and the results of these analyses follow.

\subsection{Data analysis}
Our first step in the search for the precursor QPO signal is to pass the data through a periodogram.
We initially use a Lomb-Scargle (LS) method \citep{lomb1976,scargle1982} for inspection of the frequency domain which will reveal any periodicity within the full time series data.

To isolate periodic features temporally, we use a short time Fourier transform (STFT) \citep[e.g.,][]{sejdic2009}.
This method uses a window centred at each time step to identify any periodicity within each bin and output a time-series power spectral density (PSD).

\subsubsection{Lomb-Scargle}
\begin{figure}
    \centering
    \includegraphics[width=\columnwidth]{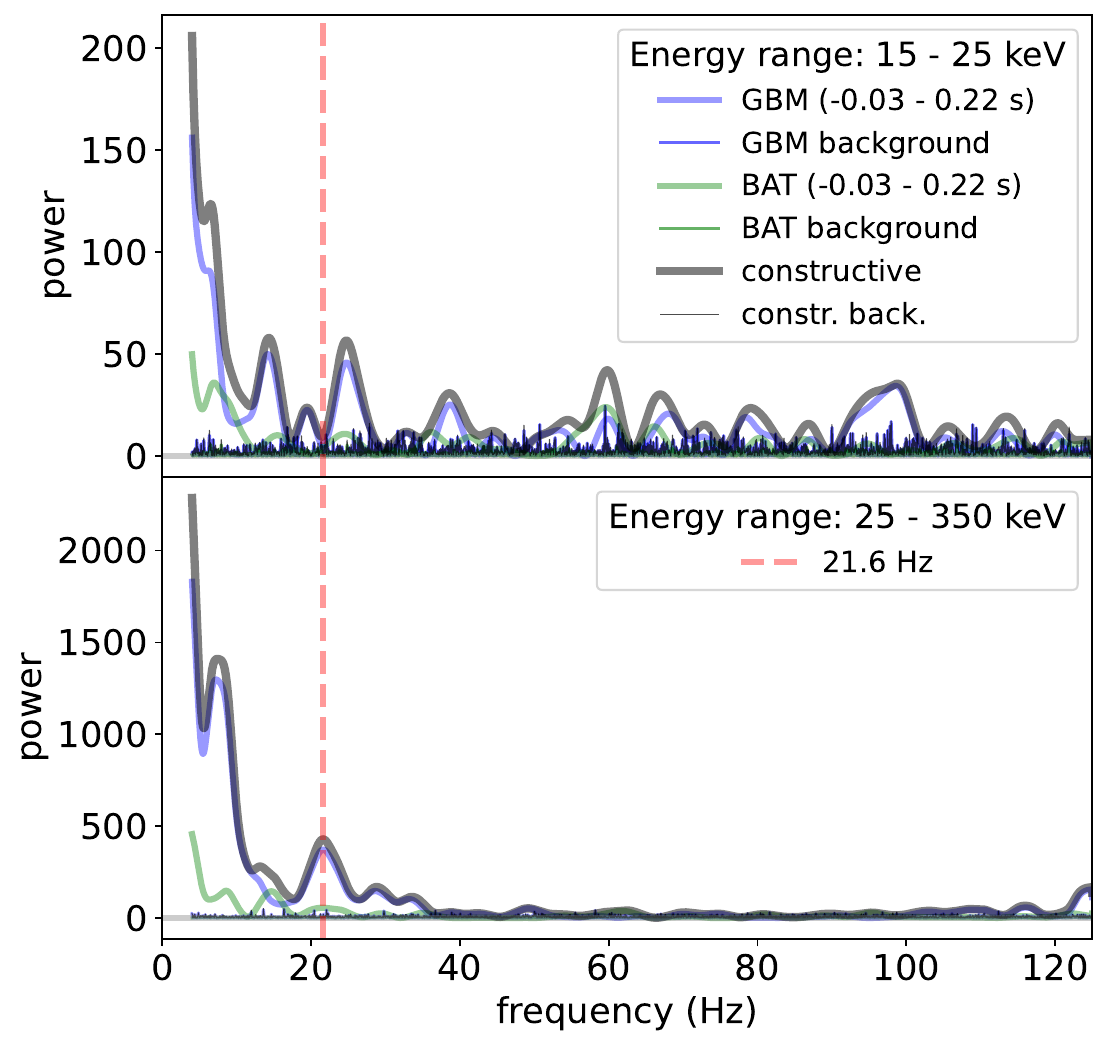}
    \caption{Lomb-Scargle periodogram of the 4\,ms binned and detrended precursor data between -0.03 and 0.22\,s of the trigger time in the low- and high-energy ranges for the \swift-BAT and the \fermi-GBM detectors. Given the sampling frequency and the duration, the frequency space is defined between $4 - 125$\,Hz. {\bf Top}: the periodogram for each detector (blue and green respectively) and the summed signal from both detectors (grey). The periodogram of the background noise from the 10s preceding the precursor is shown with a fine solid line. A 21.6\,Hz signal is marked with a red dashed vertical line. No signal at 21.6\,Hz is seen in the low energy data. {\bf Bottom}: the individual periodogram for each detector and the summed signal from both detectors. The 21.6\,Hz signal is marked with a red dashed line. Background noise level from the preceding 10\,s is shown.}
    \label{fig:LSHS_low}
\end{figure}

We limit the data to the precursor within a window of -0.03 to 0.22\,s either side of the \fermi-GBM trigger time.
The selected data are detrended using a constant for the 0.25\,s window of the precursor\footnote{For the short duration precursor we found the constant detrend via {\sc SciPy.signal} returned the same results for the \fermi-GBM data as using the {\sc GDT-core} tools to remove the background trend, based on a $\gtrsim 20$\,s interval of background data before the burst trigger time. The same constant detrend method was used with the \swift-BAT precursor data, returning a signal with zero counts line consistent with the \fermi-GBM data.}.
To ensure the time window is consistent between the instruments, BAT and GBM, we over-plot the light curves and adjust the zero-time on the BAT data to match the bin-centroid time stamps of the \fermi-GBM data.
The LS periodogram for the `low' and `high' bands are shown in Figure \,\ref{fig:LSHS_low}.

\subsubsection{Short-Time Fourier Transform}
\begin{figure}
    \includegraphics[width=\columnwidth]{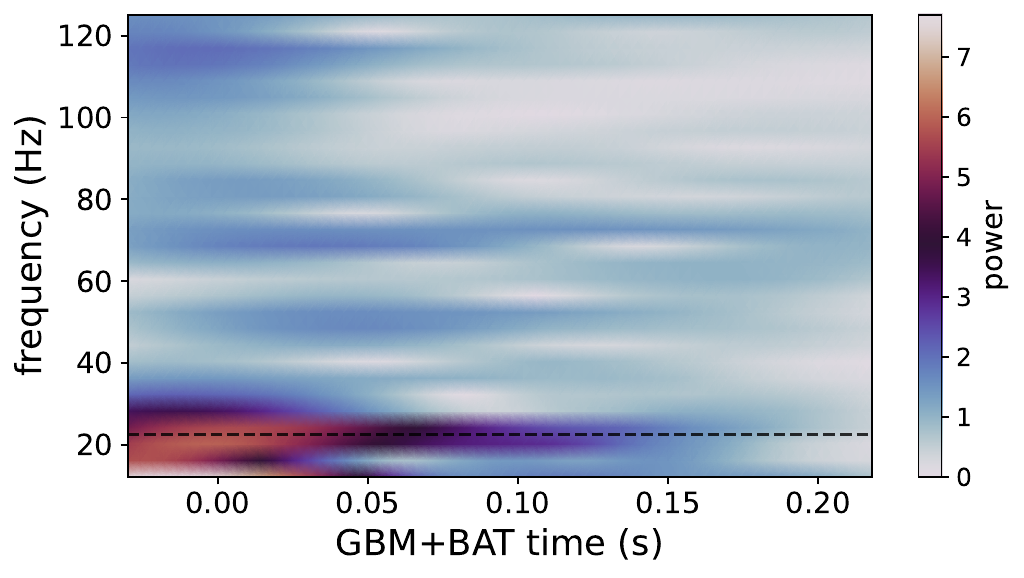}
    \caption{The combined \fermi-GBM and the \swift-BAT signal for the `all' data analysed between $10-125$\,Hz via an STFT. The colour-map indicates the relative power density at the given time. The dashed black line indicates the location of the quasi-periodic 22.5\,Hz signal from \citet{xiao2024}.}
    \label{fig:FERMI_v_BAT}
\end{figure}

To produce an STFT we start by ``Hanning a signal'' at each bin centroid \citep{blackman1958}, 
then use a Fourier transform to produce a PSD for each step.
As the windowed data extends to include the full precursor time-series, for each step, there is an $n-1$ overlap for each window function.
Due to the short duration of the precursor, this ensures that there are sufficient data within each interval for a Fourier transform, and the window function biases the signal to the periodicity in the signal for the data centred on each time step.
Using the full series will `smear' out any apparent periodicity, especially at times where the data are at $\sim$ the background level and the window function includes significant signal from the brightest precursor period, $\sim$0.0 - 0.1\,s.
The total signal for the `all' band from \swift-BAT and \fermi-GBM is shown in Figure\,\ref{fig:FERMI_v_BAT} -- the detrended and combined signal is passed through an STFT and the resulting spectrogram is plotted.
The candidate 22.5\,Hz frequency \citep{xiao2024} is indicated via a horizontal dashed black line.

\begin{figure*}
    \centering
    \includegraphics[width=\textwidth]{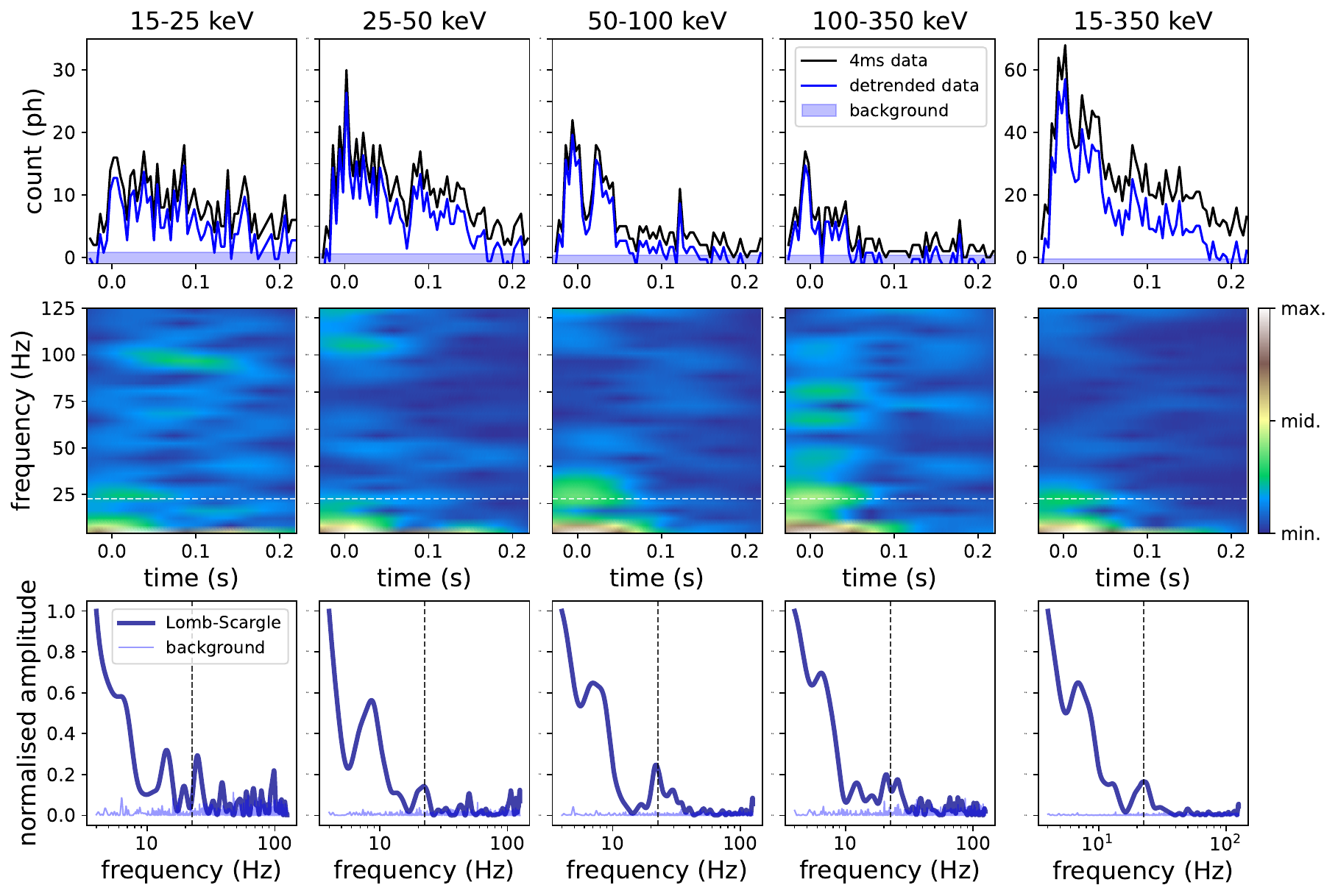}
    \caption{{\bf Top row}: light curves at band 1, 2, 3, 4, and all binned at 4\,ms from the \fermi-GBM observations of the pre-cursor, $-0.03 - 0.22$\,s. Raw data shown with a black line, and the detrended data are shown in blue.
    {\bf Middle row}: The STFT for each band over the precursor duration. The colour-scale corresponds to the amplitude of the signal at a given frequency and time relative to the amplitude range for each band. The horizontal dashed white line indicates the candidate QPO at 22.5\,Hz.
    {\bf Bottom row}: The Lomb-Scargle periodogram at each band for the series shown in the top panel. Amplitudes are normalised individually. The position of the proposed 22.5\,Hz quasi-periodic signal is shown with a dashed black line. The background signal level from the preceding 10\,s is shown as a fine solid line.}
    \label{fig:FERMI}
\end{figure*}

\begin{figure*}
    \centering
    \includegraphics[width=\textwidth]{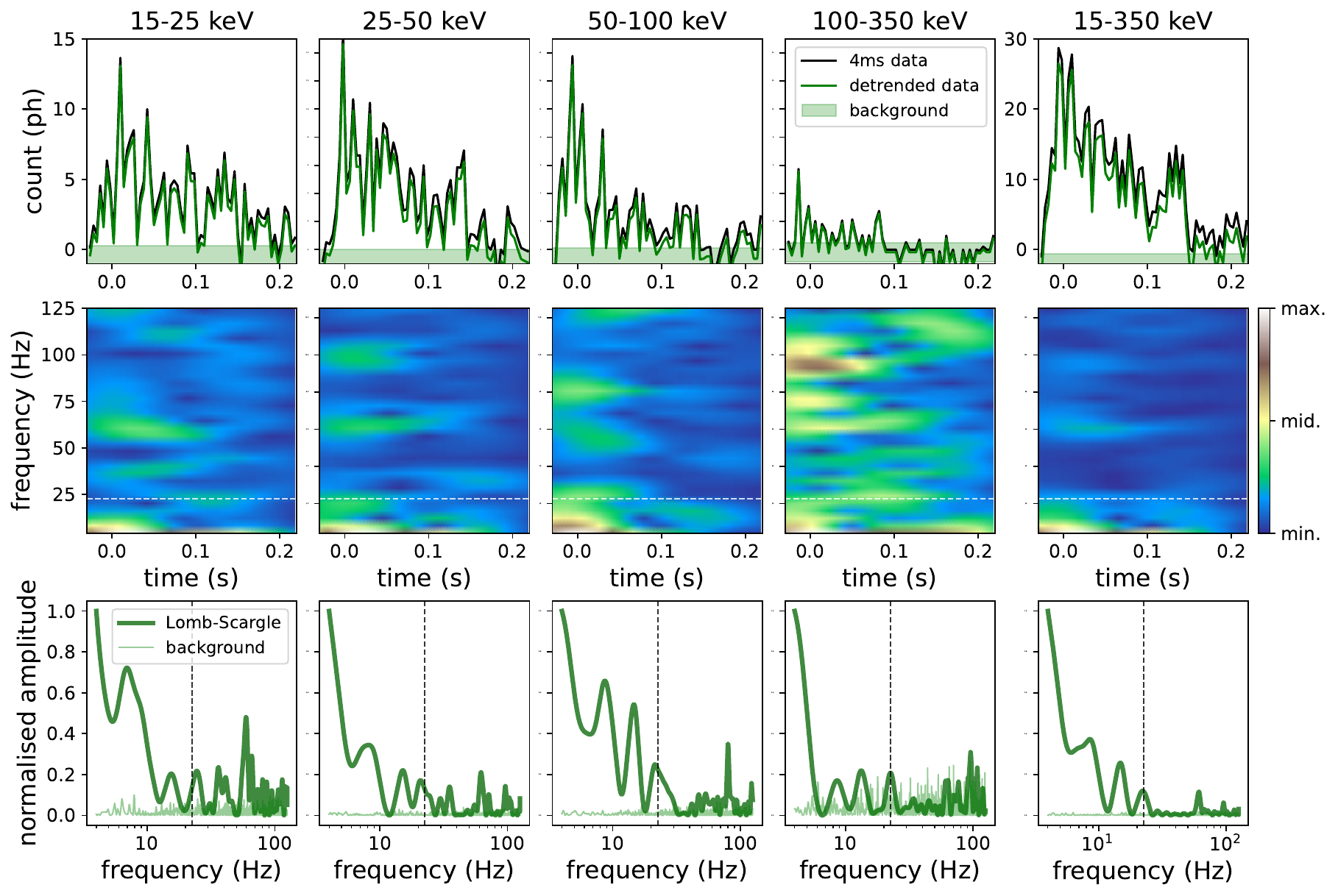}
    \caption{{\bf Top row}: Same as Figure\,\ref{fig:FERMI} but for the \swift-BAT data.
    }
    \label{fig:BAT}
\end{figure*}
Short-time Fourier transforms for band 1, 2, 3, 4, and `all' are applied individually to the light curves from \fermi-GBM and \swift-BAT.
The signal, raw (black) and detrended (blue), spectrogram, and a Lomb-Scargle periodogram for each band are shown in Figures\,\ref{fig:FERMI} \& \ref{fig:BAT} for GBM and BAT respectively.

\subsection{Periodogram Results}
Figure \ref{fig:LSHS_low} shows both the `low' (top panel) and `high' (bottom panel) data periodograms via a Lomb-Scargle.
The precursor data between $-0.03$ and $0.22$\,s are shown via a thick solid line, where blue is the \fermi-GBM, green is \swift-BAT, and grey is the combined signal.
The background noise level via a periodogram of the $-10$ to $-5$\,s data are shown as fine solid lines -- the precursor signal is seen clearly above the background levels.
The `low' data has a smaller power than that of the `high' -- in both cases, the highest signal power is seen at frequencies $\lesssim 30$\,Hz.
The low-frequency signal peaks at $\sim4$\,Hz, which is coincident with the inverse of the precursor duration, $t_{\rm dur}\sim0.2 - 0.25$\,s.
A further look at the light-curve and LS periodogram for the precursor is included in the Appendix \ref{app:A}, Figure\,\ref{fig:APP_A1}.
The LS periodogram of the `high' data reveals a signal at $21.6$\,Hz.
A power spectral density via Welch's method for \fermi-GBM and \swift-BAT data are shown in Appendix \ref{app:A}, Figure\,\ref{fig:welch}, where the QPO signal and several other periodic features are identified and indicated via vertical red lines/shaded boxes.

The short-time Fourier transform for the `all' data, shown in Figure\,\ref{fig:FERMI_v_BAT}, indicates a $\sim 22.5$\,Hz feature where a stronger signal is seen for the time period $-0.05 \lesssim t - t_0 \lesssim 0.15$\,s.
The STFT for \fermi-GBM and \swift-BAT data for bands 1, 2, 3, 4, and `all' are shown in Figures\,\ref{fig:FERMI} \& \ref{fig:BAT}.
The top row of each figure shows the individual precursor light curves for the raw and detrended data; the background noise level is indicated by a shaded region which encompasses the mean and plus/minus standard deviation for the 10\,s data preceding the precursor.
The middle row contains the STFT for each band, the colour-bar on the right indicates the relative power at each time step and frequency.
The bottom row is the LS for each band with the noise signal from the background data preceding the precursor -- the location of the QPO at 22.5\,Hz is shown with a black vertical dashed line in the LS panels, and a white horizontal dashed line in the STFT panels.
The data in band 1 do not have a peak coincident with the proposed QPO.
The STFT and LS panels show that the power is dominant at low-frequencies, $<10$\,Hz -- this is due to the precursor duration as discussed above.

Using the `high' band data periodogram in Figure\,\ref{fig:LSHS_low}, the fraction of the total precursor power in the QPO signal between $18 - 26$\,Hz is 13.2\%.
For this fraction, we assume that all of the low-frequency data are due to the precursor duration and fractional periods, while any red noise is insignificant -- see discussion in Appendix \ref{app:A}.

\subsection{Pulse profiles}
Individual GRB pulses can be modelled following \cite{norris2005}:
\begin{equation}
    N(t) = \frac{A\lambda}{e^{(\tau_1/t + t/\tau_2)}} \hspace{0.5cm} t>0,
    \label{eq:pulse_norris}
\end{equation}
where $t$ is the time after a zero time, $t_0$, and $A$ is the amplitude of the pulse; $\lambda = e^{(2\mu)}$ with $\mu = \sqrt{\tau_1/\tau_2}$, and $\tau_1$ and $\tau_2$ define the exponential rise and decay timescales of the pulse.

The photon count data (see the top panels in Figure\,\ref{fig:FERMI} \& \ref{fig:BAT}) exhibit multiple pulses within the $\Delta t \sim 0.25$\,s duration of the precursor.
\cite{xiao2024} gave the minimum variability timescale for the precursor of $\tau_m = 15 \pm 2$\,ms, which is longer by a factor of $\sim2$ than the minimum variability timescale of the main burst complex, $\tau_m = 8.5 \pm 0.8$\,ms, which is interpreted by \cite{xiao2024} as evidence for the precursor's different origin from the main burst.

To model the $\sim 0.2$\,s precursor, we sample the data within a $0.25$\,s window and use one of three pulse profile models:
\begin{enumerate}
    \item {\bf Single pulse:} A singular pulse given by eq.\,\ref{eq:pulse_norris}.
    \item {\bf Pulse + sinusoid:} A singular pulse given by  eq.\,\ref{eq:pulse_norris} and sinusoidal oscillation given by:
    \begin{equation}
        N_{\omega}(t) = \frac{A}{10} \sin\left(2{\rm \pi} \nu_{\omega} [t - x/\nu_{\omega}]\right),
    \end{equation}
    where $\nu_{\omega}$ is the quasi-periodic oscillator frequency, and $x$ is a variable between $0 - 1$. 
    \item {\bf Multipulse:} Pulses that have a profile that follows eq.\,\ref{eq:pulse_norris} but with a periodic occurrence frequency $\nu_\omega$; the number of pulses within the profile is given by the integer $\mathbb{Z} = 0.25 \nu_\omega$, and the zero-point for each pulse is $t_0(\mathbb{Z}>1) = t + \tau_2 - \left[\mathbb{Z}-1\right]/\nu_\omega$.
    The amplitude, $A$, for the pulses is given by:
    \begin{equation}
        A e^{-6\pi x \left(\mathbb{Z}-1\right)/\nu_\omega},\hspace{0.5cm}0\leq x \leq 1.
        \label{eq:pulse_amp}
    \end{equation}
\end{enumerate}
We fit these models to the summed \fermi-GBM and \swift-BAT data in the `high' band.
This energy range is motivated by the consistency between the Lomb-Scargle periodograms for our low- and high-energy bands, and the STFT of the individual bands, both indicating that the candidate QPO is only seen at energies $\gtrsim 25$\,keV (see Figures\,\ref{fig:LSHS_low},\,\ref{fig:FERMI} \& \ref{fig:BAT}).

The combined data in the `high' range are time-normalised to the peak time, $t_p$, and given a zero-point $t_0 = t_p - 0.02$\,s.
As any negative counts from Model (ii) are unphysical, and because the pulse models (i) and (iii) can only predict positive counts, we adjust the counts level by the minimum of the background, ensuring that all have a positive value.
The background in gamma-ray observations has multiple and complex potential components \citep{biltzinger2020}; as we are focused on a very short time window, the background has been determined using the data before the precursor event -- this shows a linear trend with small fluctuations $\pm 2$\,photons per 4\,ms bin.
As the precursor signal drops to levels comparable to the background within $0.2 - 0.25$\,s, the positive offset allows all counts within the $0.25$\,s duration of the precursor data to contribute to the fit for all models.
To fit models (i) -- (iii) to the data, we use {\sc Nessai} \citep{williams2021, williams2023}, a nested sampling algorithm for Bayesian inference that incorporates normalising flows.
We marginalise over the fit parameters in the ranges listed in Table\,\ref{tab:priors}.
We fix $A = 73.3$ for consistency with the peak of the combined data  -- for Model (ii), the fixed model amplitude, $A$, is reduced as the oscillation signal can contribute an additional 10\% to the peak counts.

\begin{table}
    \centering
    \caption{Prior ranges for precursor pulse models (i) -- (iii).}
    \begin{tabular}{c c c c c}
	\hline
	Model & $\tau_1$ (ms) & $\tau_2$ (ms) & $x$ & $\nu_\omega$ (Hz) \\
        \hline
	(i) & 8 -- 300 & 8 -- 300 & -- & -- \\
	(ii) & 8 -- 300 & 8 -- 300 & 0 -- 1 & 20.0 -- 30.0 \\
	(iii) & 8 -- 300 & 8 -- 300 & 0 -- 1 & 20.0 -- 30.0 \\
        \hline
        Distribution & $\log$ & $\log$ & linear & linear \\
        \hline
    \end{tabular}
    \label{tab:priors}
\end{table}

The posterior distributions from the fits for the three pulse models to the precursor data are shown in the top panel of Figure\,\ref{fig:pulse_fit}, and the corner plots for each are in Appendix\,\ref{app:B}.
We plot $400$ randomly selected samples from each model posterior.
Model (i) is shown in green; Model (ii) in orange; and Model (iii) in purple.
The horizontal dashed teal line indicates the background trend level and the shaded region the background uncertainty.

The middle panel of Figure\,\ref{fig:pulse_fit} shows the difference between the model counts and the data.
The black dashed line indicates the data - model equality, and the uncertainty due to Poisson noise and background fluctuations are shown as a grey shaded region.
Each of the models (i -- iii) are shown with the same colours as in the top panel.

The posterior parameters' median and 1$\sigma$ credible interval are listed in Table\,\ref{tab:posterior}.
We fit these models using a narrow prior for the QPO, motivated by the periodogram analysis showing a consistent signal across two instruments at 22.5 Hz and no significant features other than those at $\lesssim 10$\,Hz, however, Model (iii) sits against the lower boundary, indicating that this model, if unconstrained by the appropriate prior, will extend beyond the nominal lower bound.
The model simplicity and fixed parameters (e.g., amplitudes, rise and decline timescales, damping rate) all introduce uncertainty and the posteriors give the `best fit' for this data versus the individual models.
The log ratio for each model, based on the Bayes factor comparison and shown in Table\,\ref{tab:posterior}, gives an indication for which of these models is preferred. However, the true model is likely to be significantly more complicated than the simplified global trends of models (i) -- (iii).

A QPO in the range $20 - 25$\,Hz will produce $n\sim5 - 6$ peaks within a $0.25$\,s precursor -- for Model (iii) in Figure\,\ref{fig:pulse_fit}, the number of peaks is $n = 5$; for Model (ii), the number of peaks is $n = 6$.

\begin{table}
    \centering
    \caption{Posterior sample median and 1$\sigma$ credible interval for the model parameters fit to the data.
    The lower section gives the $\log \mathcal{B}$, of the $\ln$ Bayes factor for the comparison between the evidences for each of the models. A positive value prefers the model listed to the left of the row, while a negative value indicates a preference for the model in the column header. The best fitting model is indicated via bold text in the row and column headers.}
    \begin{tabular}{c c c c c }
        \hline
        Model & $\tau_1$ (ms) & $\tau_2$ (ms) & $x$ & $\nu_\omega$ (Hz)\\
        \hline
        (i) & $10^{+1}_{-1}$ & $56^{+2}_{-2}$ & -- & -- \\
        (ii) & $13^{+1}_{-1}$ & $55^{+2}_{-2}$ & $0.52^{+0.06}_{-0.06}$ & $24.29^{+0.60}_{-0.45}$ \\ 
        {\bf(iii)} & $\bf26^{+3}_{-2}$ & $\bf15^{+0}_{-1}$ & $\bf0.56^{+0.02}_{-0.02}$ & $\bf20.13^{+0.14}_{-0.08}$ \\
        \hline
        $\log \mathcal{B}$ & (i) & (ii) & {(iii)} & \\
        \hline
        (i) & 0 & -9.248 & -22.357 & \\
        (ii) & 0.248 & 0 & -13.109 & \\
        {\bf (iii)} & {\bf22.257} & {\bf13.109} & {\bf0} & \\
        \hline
    \end{tabular}
    \label{tab:posterior}
\end{table}

\subsubsection{Peaks and troughs}
Individual pulses within the precursor can be identified following \cite{li1996}.
Candidate peaks are found by isolating data points that are greater than both the preceding and the following data.
However, this method will identify spurious data due to noise; to mitigate this, we identify peak counts that meet the condition:
\begin{equation}
   n_p - N_v \sqrt{n_p} \geq n_{1,2},
   \label{eq:peaks}
\end{equation}
with $n_i$ being the counts at: $(i=p)$ peak,  $(i=1)$ the data point before the pulse peak, and  $(i=2)$ the data point after peak; $N_v$ is a factor in the range $3 \leq N_v \leq 5$ and accounts for noise within the data. 
Once the peaks, $n_p$, are located in this manner, then the troughs can be identified by finding the minimum counts between consecutive peaks.

\begin{figure}
    \includegraphics[width=\columnwidth]{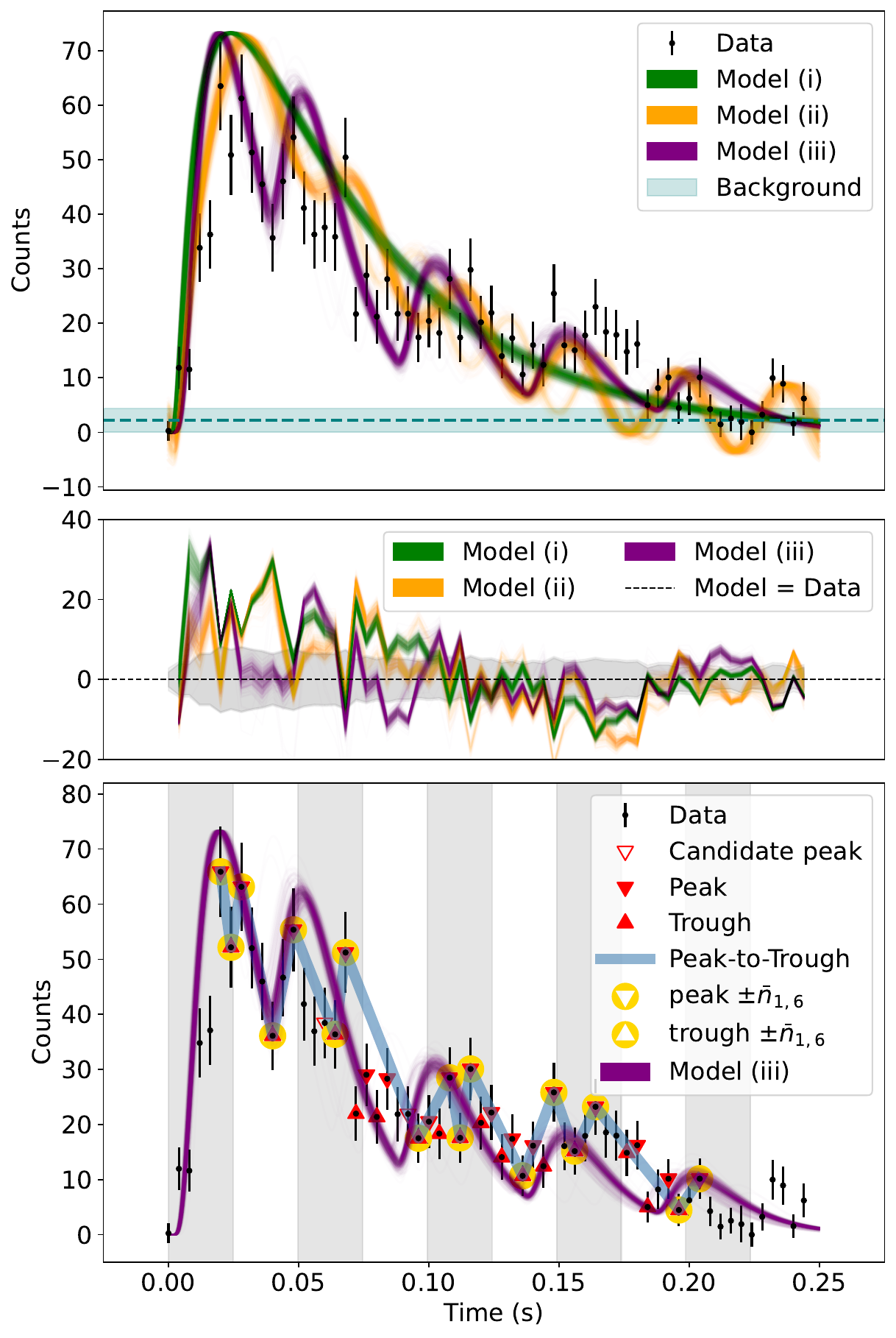}
    \centering
    \caption{{\bf Top:} The combined \fermi-GBM and \swift-BAT GRB background-subtracted, `high' band data, normalised by the peak time and offset by $0.02$\,s to give a zero-point for the start of the pulse. Three models fit to the data: green is Model (i) -- the posterior sample for a single pulse profile ; orange is Model (ii) -- the posterior sample for a single pulse plus a sinusoidal signal, ; and purple is Model (iii) -- a posterior sample from a multi-pulse model with periodic timing.
    {\bf Middle:} The difference between the data and the model. The black dashed line is a difference of zero. The grey shaded region indicates the range of the data's error bars, and the combination of the background noise and data uncertainty. The difference line for each model is shown with the corresponding model colour.
    {\bf Bottom:} The peaks and troughs identified via eq.\,\ref{eq:peaks} (red down, up triangles, with two early candidate peaks identified with hollow triangles), the pulse trend identified via the peaks and troughs (slate blue), and Model (iii) overlaid (purple). Grey vertical panels indicate half the QPO period -- equivalent to the engine duty cycle.}
    \label{fig:pulse_fit}
\end{figure}

The bottom panel of Figure\,\ref{fig:pulse_fit} shows peaks and troughs in the `high' band light curve. 
There are 17 peaks identified, giving an average variability timescale of $14.7 \pm 2.0$\,ms over the 0.25\,s duration, which is consistent with the minimum variability timescale (MVT) found by \cite{xiao2024} at $15\pm 2$\,ms.
However, the MVT between the precursor peaks is $8$\,ms, twice the bin size and therefore the smallest peak-to-peak timescale possible with this data set.
This short peak-to-peak timescale highlights the challenges in peak and trough identification, where it is likely that we have overestimated the number of true peaks.
By averaging a number of data points preceding and trailing a candidate peak, an improved estimate can be determined.
By averaging six data points each side, we find peaks and troughs with gold circles in Figure\,\ref{fig:pulse_fit}.
The number of gold circle peaks is 9, which is fewer than the simplest peak finding method but still more than found via fitting QPO models to the data.
As a candidate QPO exists within the raw data, the variability that results in the higher number of peaks in the peak-to-peak analysis, beyond any likely statistical variations (e.g., the complex at $\sim 150$\,ms), is indicative of intrinsic variability related to the emission mechanism of GRBs. 

Although significant \citep{xiao2024}, the precursor has a lower signal-to-noise ratio than the data used in \cite{li1996} to formulate the peak-to-trough identification algorithm (eq.\,\ref{eq:peaks}), which may limit the effectiveness of the unique peak identification here.
To test the effectiveness of the peak-to-tough algorithm, we generate a comparable FRED pulse with a background $\pm2.5$\,counts and Gaussian counting noise, then fit our algorithm.
The original \citet{li1996} peak-to-trough finds 15 peaks, mostly clustered around the late tail, with 3 peaks between $t_0 - 0.1$\,s. There are a further 4 candidate peaks in this same time window.
Our modified method, averaging counts for 6 points before and after, returns 6 peaks -- 2 at the maximum of the pulse at $<0.05$\,s, and 4 in the tail at $>0.13$\,s.
No periodicity is apparent in the yellow peaks for our fake pulse; -- we therefore consider our later estimate of 9 peaks within 5 pulses to be more indicative of the true variability above noise within the precursor.

\section{Discussion}\label{s:disc}
The peaks and troughs identified in the bottom panel of Figure\,\ref{fig:pulse_fit} provide an insight into the variability of the signal.
The inclusion of the noise term weakens the identification of peaks when one or both neighbouring data points are similar in amplitude. 
We average six data points before and after a candidate peak, and use eq.\,\ref{eq:peaks}.
The number of identified peaks then changes from 17 to 9, and the median variability timescale between these peaks is $\sim15\pm2$\,ms. 
The MVT from the peak-finding method remains at $8$\,ms, the minimum set by twice the time bin duration. 
Such a short MVT implies a high Lorentz factor via pair opacity arguments, with an 8\,ms timescale putting a lower limit at $\Gamma \gtrsim 80$ where we assume all of the kinetic energy is radiated as $\gamma-$rays -- an unrealistic 100\% efficiency within the precursor.

\subsubsection{Internal shock model}
Where an individual pulse is assumed to be the result of relativistic shell collisions \citep[e.g.,][]{kobayashi1997}, two shells are required to produce a single pulse.
Each shell has a characteristic mass and Lorentz factor -- we label these $M_i$ and $\Gamma_i$ respectively.
A collision between a fast and a slow shell, as a result of the different shell velocities, results in a merged shell.
If we assume an inelastic collision, then the resulting Lorentz factor will be
\begin{equation}
    \Gamma_m = \sqrt{\Gamma_f\Gamma_s\left[\frac{\Gamma_f M_f + \Gamma_s M_s}{\Gamma_s M_f + \Gamma_f M_s}\right]},
    \label{eq:merged_gamma}
\end{equation}
where the subscripts $m$, $f$, and $s$ indicate the merged, fast, and slow shell.

The number of discrete peaks for a burst complex is equivalent to one less than the minimum number of shells, or over-densities, required to produce the signal via internal collisions: $n_{\rm shells}\geq n + 1$.
For shell collisions to occur in this model, the leading shells will be slower than the trailing shells, where $ \tilde{t_i}$ is the ejection time for each shell $(i = 1...n_{\rm shells})$; then, $\tilde{t}_{n_{\rm shells}} = 0$ (i.e., all earlier shells have $\tilde{t_i} < 0$).
The width of each shell, $l_i$, corresponds to the duration of each engine activity episode as $l_i/c$ and the size of the inner engine must be $<\min l_i$.
The distance between the shells is given by $L_i = R_i - R_{i+1} - l_{i+1}$, where $R_i = -\tilde{t_i}c(1-\Gamma_i^{-2})^{1/2}$, the radius of the initial, inner edge of a shell.
Then, $L_i/c$ is equivalent to the duration of the engine's zero or low activity.

We assume that for an individual shell, the rise time corresponds to the reverse shock crossing timescale of the fast (catching) shell following a collision, where the timescale is $\tau/(1 + z)$.  The size of the shell is therefore:
\begin{equation}
    l_{i+1} = \frac{\tau c}{(1 + z)} (\beta_{i+1} - \beta_{rs}),
    \label{eq:shell-length}
\end{equation}
where $\tau$ corresponds to $\tau_1$ in eq.\,\ref{eq:pulse_norris}, $\beta = \sqrt{(1 - \Gamma^{-2})}$, and the subscript `rs' refers to the reverse shock quantity.
The reverse shock properties can be found from the merger of two shells, where the Lorentz factor of the merged shells is given by eq.\,\ref{eq:merged_gamma}, with $M_s, \Gamma_s \equiv M_i, \Gamma_i$ and $M_f, \Gamma_f \equiv M_{i+1}, \Gamma_{i+1}$.
Then, the reverse shock Lorentz factor is \citep{kobayashi1997}:
\begin{equation}
    \Gamma_{rs} = \sqrt{\frac{1+2\Gamma_m/\Gamma_{i+1}}{2+\Gamma_m/\Gamma_{i+1}}},
    \label{eq:rsgamma}
\end{equation}
leading to $\beta_{rs} = \sqrt{1-\Gamma_{rs}^{-2}}$.

The MVT, $\tau_m$, of the prompt emission provides a lower limit on the Lorentz factor: $\Gamma \gtrsim 80 (E/10^{51}{\rm erg})^{1/5}(\tau_m/0.1 {\rm s})^{-2/5}$, where $E$ is the jet isotropic equivalent total energy \citep{lamb2016}.
The precursor has an isotropic $\gamma$-ray energy $E\sim 6.97\times 10^{48}$\,erg \citep{xiao2024} -- however, the energy in $\gamma$-rays is only a small fraction of the total energy.
The multiple internal shock collision model is classically found to result in low efficiencies, $< 0.1$ \citep{mochkovitch1995}, however, see \cite{kobayashi2001} for the case of ultra-efficient internal shocks.
For an efficiency of $\sim 0.02$, the total jet energy would be on the order $E_{\rm total} \sim 3.5\times 10^{50}$\,erg, and using $\tau_m/(1 + z)=14(7.4)$\,ms, the Lorentz factor is $\Gamma > 142(175)$.
A low efficiency of $\eta\sim 0.02$ implies that the range of Lorentz factors across the precursor outflow is a factor $\sim2$ \citep{kobayashi1997}.

It is not trivial to analytically estimate the parameters for a minimum of 10 shells colliding in series to give the resultant pulse profile seen with Model (iii).
We therefore use the number of unique pulses identified by the the peaks-to-trough algorithm, using $n = 9$ (yellow points in Figure\,\ref{fig:pulse_fit}).
The QPO is then the result of a periodic jet on/off timescale wherein the `on' cycle ejection of impulsive shells is stochastic.
We note that there are $\sim2$ peaks within each of the pulses of Model (iii) for the light curve, suggesting an average of 3 shells ejected per `on' cycle.

The energy per pulse is proportional to the number of photons, and each pulse has an identical profile as eq.\,\ref{eq:pulse_norris} in terms of the parameters $\tau_1$ and $\tau_2$ (thus ensuring the periodicity) but a unique amplitude, $A_i$; then, the energy per pulse scales as $E_i = E_{\rm total} A_i/\sum_iA_i$.
Each pulse has $n\geq 2$ shells, and the energy for each shell is identical such that $E_i/n = S_i$, where $i = 1,\dots n$.
Each of the shells within an engine cycle will collide and merge within the half-period, $P/2 \sim 25$\,ms,  of the on/off cycle.
As the observed $\gamma$-ray emission is only produced when shells collide, the assumed number of shells per pulse is the minimum; in reality, there can be many more components that do not collide in this period or which carry insufficient energy to create a bright pulse.
Each pulse complex preserves the periodicity of the engine activity despite the stochastic nature of the unique pulse shells -- this is evidenced in the alignment of identified peaks with the grey `on' cycles shown in the bottom panel of Figure\,\ref{fig:pulse_fit}.

We sub-divide the peaks and troughs into distinct interval groups.
Each group has a temporal separation between neighbouring peaks of $\Delta t\geq 20$\,ms.
The peaks naturally split into 5 clusters of $n \sim 2$ peaks.
We use these to infer the various energy, mass, and Lorentz factor requirements to recreate the observed precursor phenomenology.
We assume that the number of shells per pulse is $n_{\rm shells} = n+1$, and that the kinetic energy within each pulse is distributed equally amongst the individual shells.
The shell minimum Lorentz factor, $\Gamma_i$, is determined in the same way as we previously found the minimum Lorentz factor of the precursor given the MVT -- however, here we use the time from the preceding trough to peak (i.e., the rise time) and the individual shell energy\footnote{For merged shells, the total energy is the sum of the two merging shells minus the energy converted to internal energy. The energy conversion efficiency for our system is on the order of $10^{-4}$ to $10^{-2}$ and so is not evident in the listed values in Table\,\ref{tab:peaks_in_pulses}.}.
The shell collision times are assumed to be the preceding trough time, the shell width is given by eq.\,\ref{eq:shell-length} and $\beta_{rs}$ is found with eq.\,\ref{eq:rsgamma}, and mass is the isotropic equivalent assuming $M_i = E_{i, {\rm shell}}/(\Gamma_i c^2)$. Quantities are listed in Table\,\ref{tab:peaks_in_pulses}.

\begin{table}
    \centering
    \caption{Individual peaks identified via the yellow markers in the bottom panel of Figure\,\ref{fig:pulse_fit} tend to line up with the half-QPO period shown with grey panels. Pulses are separated by $24.7/(1 + z)$\,ms, and individual peaks correspond to the collision of two shells -- thus, each pulse has $n+1$ shells. We assume the initial shell within each pulse is the first shell collision product for each pulse. The initial Lorentz factors are based on the minimum $\Gamma$, given the shell energy and the timescale from previous trough to peak. 
    All timescales are in the rest frame.}
    \begin{tabular}{c | c c c c c c}
        \hline
        Pulse & $n_{\rm peak}$ & $t_{0,\rm collision}$ & $\Gamma$ & $E_{\rm shell}$ & $M_{\rm shell}$ & $l/L$ \\
            &       &   ms    &     & $\times10^{49}$\,erg & $\times10^{-7}$\,M$_{\sun}$ & \\ 
        \hline
        1   & 0+1 &   0 & 86 & 10.0  & 6.48  & -- \\
            & 2 &   22  &   163   &   5.0  &   1.70  & 0.63  \\
        2   &   0+1   &   37  &   112   &   6.0  &   3.02 &    0.60 \\
            &   2   & 59 & 148 & 3.0  &  1.14 & 0.14  \\
        3   & 0+1 &   89  &   87    &   3.7   &   2.38   &   0.39  \\
            & 2 &   104 &   134   &   1.8 &   0.78    &   0.57    \\
        4   & 0+1 &   126 &   78    &   2.2 & 1.60   &   0.53  \\
            & 2 &   145 &   92   &   1.1 & 0.68 &   0.80    \\
        5   & 0+1 & 182 &  90    &  2.0 & 1.26  &   0.22  \\
        \hline
    \end{tabular}
    \label{tab:peaks_in_pulses}
\end{table}

For each pulse, the first peak is the result of the collision of shells 0 \& 1 within that pulse complex.
The individual Lorentz factors for each shell, across all pulse complexes, have values such that no shell from an earlier pulse will be caught by the merged shell (eq. \ref{eq:merged_gamma}) in the following pulse.

The decaying peak luminosity structure is a distinct feature of the precursor, and may be related to the damping of the mechanism responsible for the periodicity.
This analysis demonstrates that the precursor's temporal phenomenology can be recreated with a simple baryonic shell collision model for GRB prompt emission.
Via the multiple shell modelling, and by summing the individual shell masses in Table \ref{tab:peaks_in_pulses}, the total mass within the precursor is $1.90\times10^{-6}$\,M$_{\sun}$, if the emission were isotropic.
Using the jet half-opening angle\footnote{The precursor outflow may have a different opening angle than that determined by the afterglow model, which we use here for convenience and lack of better information.} $\theta=3{\fdg}27$ from the afterglow modelling in \cite{rastinejad2022} and assuming bipolar jets, the mass in the precursor is $2.99\times10^{-9}$\,M$_{\sun}$; assuming a jet-to-accretion efficiency of $\eta\sim0.01$, the accretion rate for the precursor would be $\simeq2.99\times10^{-7}$\,${{\rm M}}_{\sun}$s$^{-1}$ -- well within the constraints expected for a low-energy jet.
If we assume an accretion rate $\sim10^{-2}{{\rm M}}_{\sun}$s$^{-1}$, then the accretion-to-jet efficiency for the precursor is $\eta\sim 10^{-5}$, consistent with the $NN-T_c$ and $NN-T_s$ models from \cite{gottlieb2023}, where the $T$ indicates an initially toroidal magnetic field.

\subsubsection{Implications of the precursor Lorentz factor}
The rapid variability within the precursor, and the apparent periodicity when the data are compared with the best fitting model (see the top and middle panel of Figure\,\ref{fig:pulse_fit}) requires a minimum Lorentz factor $\Gamma \gtrsim 100$.
Such a rapid outflow is inconsistent with precursor emission as a result of shock or cocoon breakout, where the light curve will be smoother and the Lorentz factor low: $\Gamma\sim2$ to $10$ \citep[e.g.,][]{wang2007, gutierrez2024}.
However, for long GRBs, precursor emission {\it has} been associated with shock or cocoon breakout \citep[e.g.,][]{ramirez-ruiz2002, nakar2017}, while short GRB precursors show little difference to the main burst complex \citep{troja2010} and are likely to have an engine origin.
Given that GRB\,211211A is associated with a candidate kilonova and defines a ``new'' class of long, merger origin GRBs \citep{gompertz2023}, the precursor emission in such merger origin systems will reflect early engine activity, as for short GRBs.

For a merger, the ejecta envelope that the jet has to ``drill'' through is significantly smaller than the case of a collapsing massive star \citep{bromberg2012}, and given the asymmetric nature of merger ejecta, the polar direction is expected to be relatively free of mass at early times -- low-density ejecta in the polar directions would barely resist jet penetration, and any energy dissipated via early activity will produce precursors phenomenologically similar to the GRB main burst complex.
The turbulent nature of early accretion following a neutron star merger provides ideal conditions for periodicity driven by instabilities within the disk that couple to a jet.

\subsection{Black hole engine case}
The engine size from the MVT is $\tau_m c \lesssim 2.4 -4.5 \times 10^8$\,cm, and $\approx 10 - 100 \times r_{\rm disk}$ for a neutron star merger.
From the peak-to-trough analysis we find variability on a 4\,ms timescale (equal to the bin size, suggesting the true variability minimum is shorter); in the rest frame this results in an engine size $\lesssim 1100$\,km, or $1.1\times 10^8$\,cm.
Using the more conservative MVT, if the emission has its origin within a jet with a bulk Lorentz factor $\Gamma\gtrsim140$, then the dissipation radius will be at $R_d \lesssim \Gamma^2 \tau_m c \sim 4.5 - 8.8\times10^{12}$\,cm, and consistent with the expectation for GRB prompt emission.

Periodicity within a jet origin precursor suggests that the jet is coupled to the precession mechanism.
In the case where the jet is powered by the Blandford-Znajek mechanism \citep{blandford1977}, a QPO from LT precession of the accretion disk will have an oscillation frequency on the order of the inverse gravitational time, $t_g = GM_\bullet/c^3$.
Where the QPO frequency is the inverse of the dynamical time $t_{\rm dyn} = \sqrt{r^3/ GM_\bullet}$, with $t_{\rm dyn} > r/c$, then the jet powering mechanism will be Blandford-Payne \citep{blandford1982}. 

\begin{figure}
    \centering
    \includegraphics[width=\columnwidth]{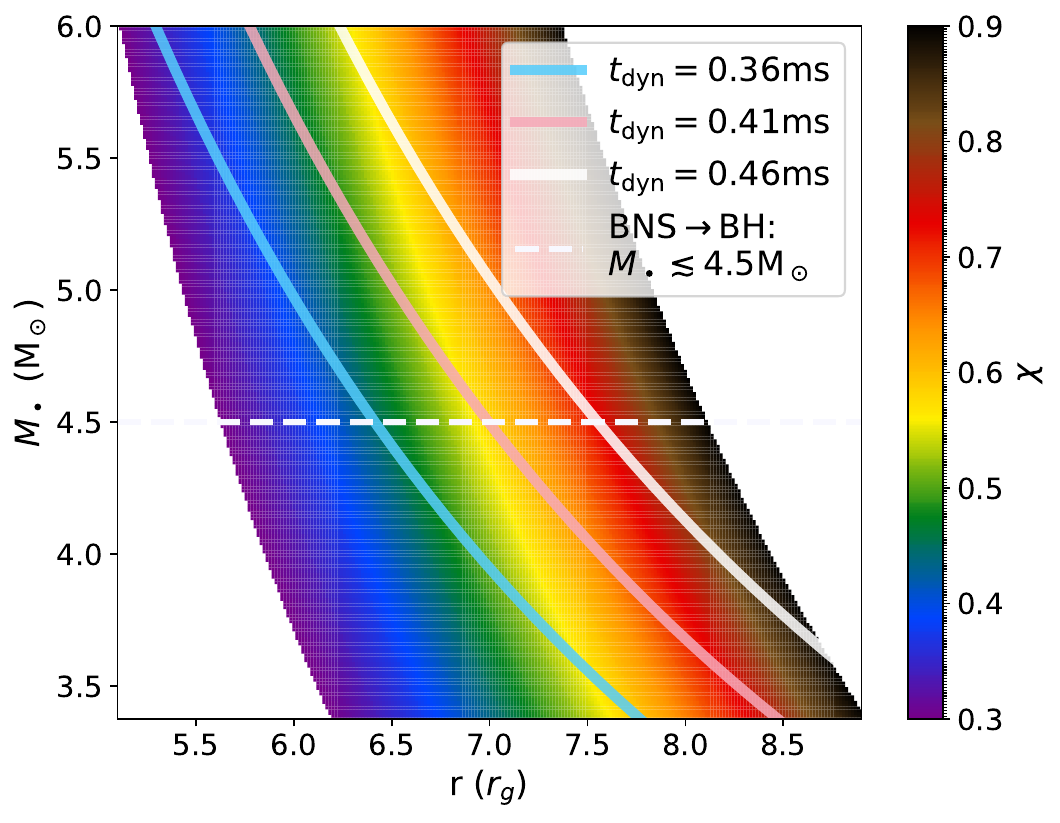}
    \caption{ The black-hole spin for Lense-Thirring precession at $22.5 (1 + z)$ Hz shown with the colour-map, for a disk radius of $5.1\leq r/r_g \leq 8.9$ as a factor of the gravitational radius, $r_g = G M_\bullet / c^2$ (and equivalent to $t_g = G M_\bullet / c^3$) and mass $3.38 < M_\bullet/{\rm M}_{\sun} < 6.00$. The solid contours are lines of constant dynamical time, $t_{\rm dyn} \simeq \sqrt{r^3/ GM_\bullet}$. The horizontal white dashed line indicates the $\sim$ maximum black hole mass following a binary neutron star merger, set at $2\times M_{\rm TOV}$.}
    \label{fig:LT_space}
\end{figure}

The mass range for the black hole engine is assumed to be between BH prompt collapse threshold \citep{sarin2021}, $M_\bullet \gtrsim (3/2)M_{\rm TOV} \sim 3.4$\,M$_{\sun}$, where $M_{\rm TOV} \simeq 2.25$ is the Tolman–Oppenheimer–Volkoff (TOV) mass \citep{yi-zhong2024}, and $M_\bullet \lesssim 4.5$\,M$_{\sun}$, though we extend the black hole mass limit to $6.0$\,M$_{\sun}$ to include black hole neutron star merger scenarios.
For the LT precession at $22.5 (1 + z)$\,Hz, and at a given radius, the required black hole spin $0.3 \leq \chi \leq 0.9$, (eq.\,\ref{eq:LTfreq}), is shown in Figure\,\ref{fig:LT_space}.
The timescale and radius for a 22.5 Hz QPO from a $\chi>0.3$ black hole in this mass range is $100 \leq t \leq 150$\,$\mu$s and $30 \leq r \leq 45$\,km, respectively.
From Figure\,\ref{fig:LT_space}, the LT effect at the QPO frequency can exist if the radius of the precession is $\sim 5$ to $9~r_g$.
This is consistent with the disk just beyond the inner-most stable circular orbit for a rotating ($\chi = (0.3,~0.9)$) black hole -- $r_{\rm ISCO} \sim (4.9,~2.3)\,r_g$. It is also potentially indicative of a disk tear due to early warping as a result of the LT precession \citep{nealon2015}.

Amongst the alternative black hole disk origins for the QPO, many have periodicity due to Keplerian motion.
Such oscillations include acoustic p-modes, orbital resonances, and blobs within the disk.
For diskoseismology, g-modes are comparable to p-modes at large radial distances (i.e., the outer disk) where $N_{BV}/\kappa \sim 1$.
For the inner radius, this ratio is typically $N_{BV}/\kappa > 1$, and the g-mode frequency will be higher than that of the p-mode, which is equivalent to the Keplerian frequency.

\begin{figure}
    \centering\includegraphics[width=\columnwidth]{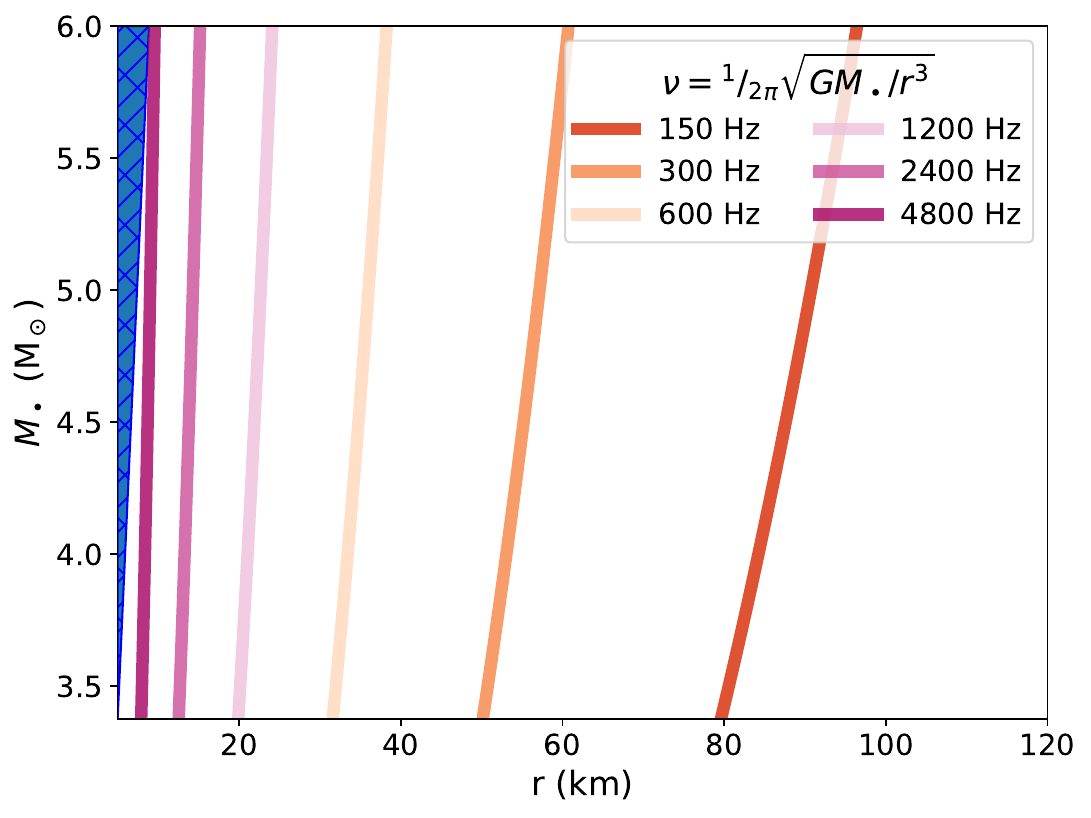}
    \caption{Given the black hole mass and radius within the disk, the QPO from the Keplerian frequency ranges from $10^2$ to several $10^{3}$\,Hz, well beyond the $22.5$\,Hz of the observed QPO. The blue shaded hatched region gives the lower limit on the radius, $r = r_g(M_\bullet)$.}
    \label{fig:keplerian}
\end{figure}

We plot the Keplerian frequency for black hole mass and disk radius of our system in Figure\,\ref{fig:keplerian}. 
The oscillation frequency $\nu_{\phi}$ of Keplerian type oscillators  scales with radius as $\propto r^{-3/2}$, with $\nu_{\phi}\sim 12$\,kHz at a radius of $r_{g}$, $\sim2$\,kHz at the ISCO, and decreasing further to $\nu_{\phi}\sim 100$\,Hz at $100$\,km (see Figure\,\ref{fig:keplerian}). 
For the epicyclic frequencies (eq.\,\ref{eq:freqepi}), using our parameter space, the radial epicyclic frequency at the observed QPO corresponds to distances around $\gtrsim 90 ~r_g \sim 590$\,km, and vertical oscillations at a radius $\gtrsim 290$\,km.
When the epicyclic frequencies form integer ratios, such as $\
{\nu_{\theta}}/{\nu_{{r}}}={3}/{2}$, resonances can amplify the oscillatory modes, producing QPOs. 
Given the mass range of the system, the resonant radius has a lower limit $r>1800$\,km, and located beyond the edge of the disk. 
Additionally, the frequency in the 3:2 resonance is below the observed QPO at $\nu_{3:2} \sim 1.9:1.3$\,Hz.
The periastron precession frequency will be at the observed QPO, $\nu_{\rm low}\sim22.5$\,Hz for a radius of $\approx 450$\,km. 
Orbital resonances at the QPO frequency take place beyond the scale of the disk system from a merger.

\begin{figure}
    \centering\includegraphics[width=\columnwidth]{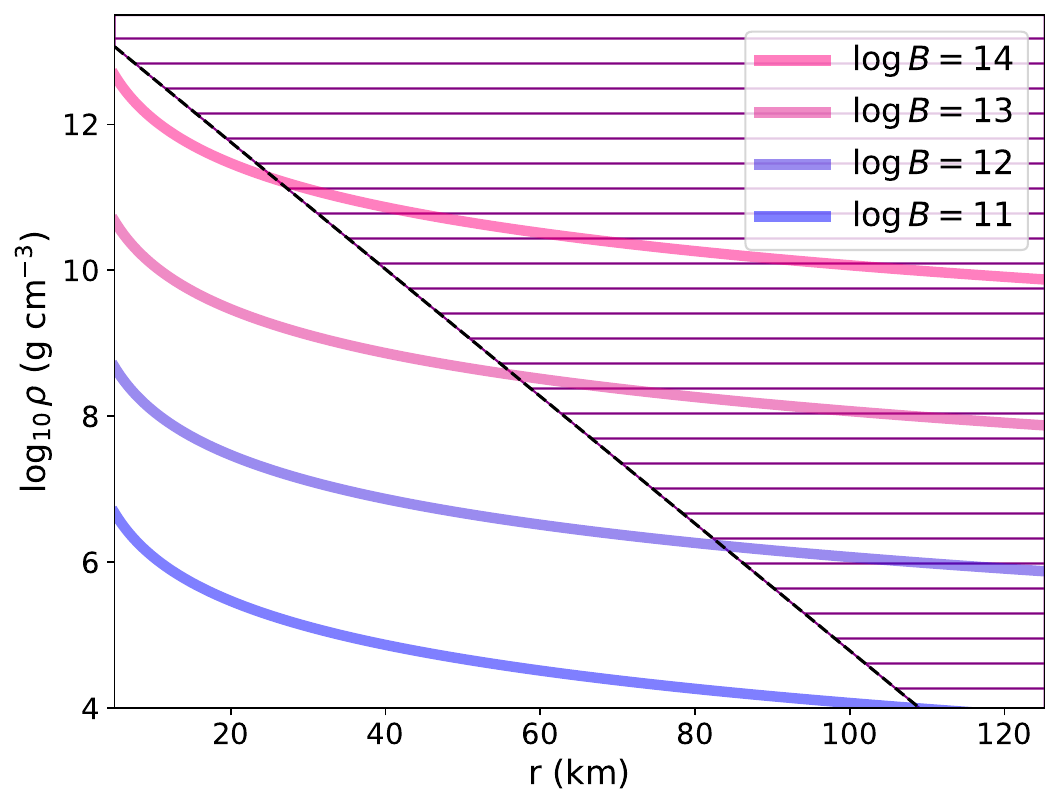}
    \caption{Parameter space and QPO within a BH disk given MHD instability scenarios. MHD or MRI oscillations, the white region indicates the potential parameter space for a (1 + z) 22.5\,Hz QPO from a black hole accretion disk. Lines of constant B-field are shown with coloured contours. The horizontal hashed region is above the expected disk density \citep{kawaguchi2025}. 
    }
    \label{fig:theothers}
\end{figure}

For MHD MRI oscillations, the frequency is shown in Figure\,\Ref{fig:theothers}.
For the MHD oscillations, based on the Alfv\'en wavelength, there is parameter space for a 22.5\,Hz QPO from the black hole system.
Figure\,\ref{fig:theothers} indicates this viable parameter space, notably the region to the lower left of the lined zone -- for a 22.5\,Hz QPO, the required magnetic field is $10^{11} \lesssim B \lesssim 10^{14}$\,G, with higher disk densities requiring a stronger magnetic field and a smaller radial distance to the instability responsible for the QPO.
An MRI within the inner disk regions of a turbulent accretion disk, the `dead zone', will result in episodic accretion giving an apparent duty cycle for the engine \citep{masada2007, masada2009}.

It is worth noting that, although we consider only a QPO within the disk, MHD instabilities can also occur at the base of, or within, the jet. 
The order of magnitude for the conditions required by such instabilities are similar to those for the inner disk case -- and require a length-scale comparable to $r_g$.

\subsection{Magnetar engine case}

The constraints on the maximum size of the engine from the rise times of the modelled pulses ($\tau_1 = ~26/ (1 + z)$\,ms for the preferred Model (iii) ) are $\tau_1 c\lesssim 7200$\,km.
This length scale is nearly two orders of magnitude larger than the expectation for a neutron star merger disk system, typically $\lesssim 200$\,km \citep[e.g.,][]{rosswog2024}.

If the observed QPO is coupled to the magnetar spin, then we can infer a spin period of $P \sim 44/(1+z)$\,ms.
Such spin-QPO coupling would place the oscillation at the co-rotation radius within the disk, $r_{c} = (GM P^2/(2\pi)^2)^{1/3} \approx 270$\,km.
Alternatively, the  Alfv\'en frequency, $\nu_A \equiv \nu_{\rm QPO}$, (assuming the Alfv\'en frequency sets the QPO frequency) within a magnetosphere places a limit on the magnetic field strength.
A limit on the maximum radial extent for Alfv\'en wave oscillations is the Alfv\'en radius, $r_{\rm A} = \mu^{4/7} (G M)^{-1/7} (3M_{\rm disk}/t_\nu)^{-2/7}$, where $\mu = B R^3$ is the magnetic moment (and constant at all radii above the surface for a dipole); $M_{\rm disk} \sim 0.1$\,M$_{\sun}$ is the mass of the disk \citep{hotokezaka2013}, and $t_\nu = r_{\rm disk}/(\alpha c_s)$, where we fix the dimensionless viscosity parameter\footnote{Estimates of the value of $\alpha$ can vary \citep[e.g.,][]{starling2004,king2013,martin2019}.} as $\alpha\sim0.1$  \citep{king2007}, and the sound speed is $c_s \leq c/\sqrt{3}$.

\begin{figure}
    \centering
    \includegraphics[width=\columnwidth]{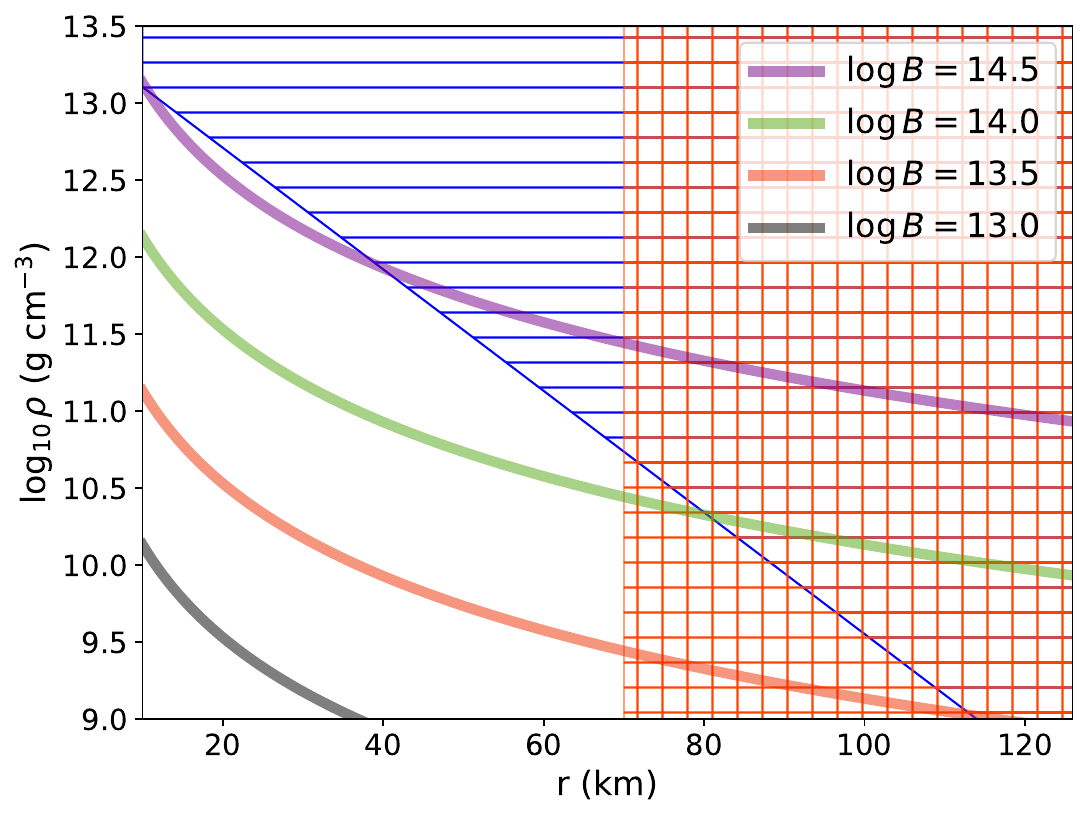}
        \caption{The magnetic field strength required for a QPO from Alfv\'en waves at $(1 + z)~22.5$\,Hz, with the radial distance (within an accretion disk) and the corresponding density (from the solution to equation\,\ref{eq:QPOR}). Coloured contours indicate dex $B-$field values as given in the legend. The orange grid region lies beyond the maximum Alfv\'en radius, and is excluded. 
        The blue striped region lies beyond the maximum density profile for the accretion disk and is also excluded.}
    \label{fig:B_Alfven}
\end{figure}

The range of the required magnetic field strength for a QPO at the observed frequency, given the constraints on the radius and density of the plasma, is shown in Figure\,\ref{fig:B_Alfven}, where the x-axis shows the range in radius, $1 < r ~(\times 10^6 {\rm ~cm)} < 32$, and the y-axis shows the range in density, $9.0 < \log\rho {~~\rm (g~cm^{-3}})< 13.5$ \citep[see e.g., Figure 1 \& 5 in, ][]{rosswog2024}.

The maximum field strength of a magnetar, $B_{\rm max}$, is determined by equating the magnetic energy with the gravitational binding energy, resulting in a maximum limit on the magnetic field strength:
\begin{equation}
    B_{\rm max} \sim \left(\frac{GM_{\rm TOV}^2}{r_m^4}\right)^{1/2}.
    \label{eq:virial}
\end{equation}
We find that our system never violates this limit.

In Figure\,\ref{fig:B_Alfven} we assume a surface radius $r_{m}\sim11$\,km, the magnetic field above the surface declines as $B\propto r^{-3}$ for a dipole field and the density profile within the disk scales with radius as $\rho(r) = \rho_0 e^{-r/r_{m}}$\,g\,cm$^{-3}$.
The dependence on the radius within the disk for a given $\nu_{\rm QPO}$ is then
\begin{equation}
    \nu_{\rm QPO} = \frac{1}{r_A(r)} \frac{B}{\sqrt{4\pi\rho_0}}\left(\frac{r}{r_{m}}\right)^{-3} e^{r/2r_{m}}.
    \label{eq:QPOR}
\end{equation}

If the QPO is coupled to the spin, the $\sim$41\,ms period is within the expected estimates for a neutron star merger remnant.
For this case, the luminosity due to dipole radiation is \citep[e.g.,][]{gibson2017}
\begin{equation}
    L_{\rm d} = \frac{\eta_d}{(1-\cos\theta)} \frac{\mu^2 \nu_s^4}{6 c^3},
    \label{eq.dipole}
\end{equation}
where $\eta_d$ is the dipole radiative efficiency, $\theta$ is the half opening angle of the emission cone or jet, and $\nu_s$ is the magnetar spin frequency.

For a QPO that has its origin in the Alfv\'en wave frequency, the spin frequency is related to the Alfv\'en frequency via a harmonic number, such that $\nu_A = H \nu_s$.
The dipole luminosity will reduce for higher harmonics, given a fixed efficiency and magnetic moment.
The observed precursor luminosity is $L = 2.53^{+0.06}_{-0.47}\times10^{49}$\,erg s$^{-1}$ and the beaming corrected luminosity is $L_j = 4.11\times10^{46}$\,erg s$^{-1}$, where we use the jet half opening angle from afterglow modelling in \citet{rastinejad2022} (note that the precursor may have a different opening angle to the jet responsible for the afterglow). 
The dipole luminosity (eq.\,\ref{eq.dipole}), beamed into a jet half opening angle of $\theta_j = 3\fdg27 = 0.057$\,rad, is $4.2\times10^{46}$\,erg s$^{-1}$, where we assume an efficiency of $\eta_d\sim0.02$, and is consistent with the precursor beaming corrected power.
However, as dipole radiation requires a rigidly rotating neutron star, and occurs on the order of $\mathcal{O}(10{\rm s})$, dipole luminosity is inconsistent with the precursor timescale.

In the magnetar origin precursor scenario, if the lifetime is $\gtrsim 10$\,s (i.e., a stable and long lived magnetar), then dipole radiation should be seen in the kilonova -- the bolometric luminosity will be higher and the light curve peak timescale marginally earlier than the case without a magnetar engine \citep[see e.g.,][]{sarin2022, ai2025}.
The dipole radiation from a magnetar is an additional term that adds to the internal energy of the merger ejecta \citep[see eq.\,9 in][]{sarin2022} and increases the total luminosity of the kilonova.
The addition of dipole radiation from a $\sim$few ms spin magnetar increases the kilonova luminosity by a factor $\sim{(10 - 1000)}$, where the spin frequency is related to the dipole power from the magnetar via $L_{\rm d} \propto \nu_s^4$.
However, as our period is approximately a factor 10 larger than a classical ms magnetar, the total dipole luminosity will be approximately 4 orders of magnitude smaller.
The power budget for injection into the ejecta is then on the order $\mathcal{O}{(10^{45}\,{\rm erg~s}^{-1})}$, and despite the efficiency for the injection of the dipole power $\xi\rightarrow 1$ at early times \citep{sarin2022}, when compared to the power from radioactive decay on the order $\mathcal{O}{(10^{49} {\rm erg~s}^{-1})}$, the spindown contribution from such a long-lived and stable magnetar is negligible.
The light curves for a $L_d = 10^{45}$\,erg s$^{-1}$ magnetar powered kilonova are shown explicitly in Figure 2 of \cite{ai2025}, where a comparison with a non-engine fed kilonova is made.
For such a long-lived magnetar engine, a late radio transient would be expected from the emergence of the pulsar wind nebula \citep{omand2018, omand2024}.
Such a radio transient would be expected at $\gtrsim100$\,days post burst -- however, such a transient remains elusive \citep{schroeder2024}.

A magnetar with a spindown power on this order will contribute insignificantly to the kilonova light curve at $1-10$\,days.
The kilonova following GRB\,211211A was shown to be inconsistent with the expectation from a magnetar-powered transient \citep{rastinejad2022} -- however, their magnetar kilonova models assumed a dipole luminosity approximately 3 orders of magnitude larger.
For a magnetar spin-coupled QPO, the oscillation radius is $\sim$300\,km, and beyond the scale expected for a merger system.
Alternatively, for magnetar cases where the QPO does not reflect the rotation frequency, then our assumption for the magnetar spin should place it at $\sim$few\,kHz \citep{sarin2022, ai2025}, which would add significantly to the kilonova on timescales ruled out by the observations, for a long-lived and stable neutron star \citep{rastinejad2022}.
Thus, for a magnetar origin of the QPO, we require a spin on the same order as the QPO frequency to ensure a low dipole luminosity, and the QPO radius to be significantly smaller than the spin co-rotation radius. 
Although we cannot rule out a magnetar origin, the required properties for such a system are unusual.

If the precursor duration reflects the timescale for a hypermassive neutron star (HMNS), then the end of the precursor and gap to the main burst complex could indicate the collapse to a black hole. 
In such a scenario, the total mass of the merger system should be $M_{\rm total} < 2.8$\,M$_{\sun}$, and the mass ratio should be $q\gtrsim1.3$ -- where we adopt the magnetically arrested disk model to produce the long duration GRB \citep[see Fig. 2 in][]{gottlieb2023}.
The required HMNS spin frequency would be $0.5 \lesssim \nu_s \lesssim 3.5$\,kHz, with a radius $\sim 20$\,km \citep{rosswog2002}, implying that if the engine is a HMNS, then the observed QPO will not be of co-rotation origin and will therefore not be tracing the remnant spin period.
Similar to the case for co-rotation above, if the HMNS rotation is not linked to the QPO frequency, then the rotation is likely on the order of a few\,kHz.
As a HMNS will collapse to a black hole on a relatively short timescale, and the collapse can occur before the rotational energy is depleted \citep{bernuzzi2020}, then no energy will be deposited within the ejecta and the kilonova will be unlikely to deviate from the non-engine fed models.
In the HMNS scenario, the QPO can be produced via Alfv\'en wave oscillations within the inner disk at 20 -- 50 km (see Figure\,\ref{fig:B_Alfven}).

A magnetar or HMNS central engine does not rule out the internal shock model presented in Table\,\ref{tab:peaks_in_pulses}.
For such an engine, the individual shells will be accelerated by the same jet launching processes as the black hole engine, where jets from accretion are ubiquitous in astrophysical systems.
The periodicity of the signal, due to Alfv\'en oscillations, will be imprinted in the engine `on' and `off' cycles.
The nature of the shell ejection within each `on' cycle could be a function of the details of acceleration in early jet launching physics and we note that the average number of peaks per `on' cycle is $\sim$constant.

For GRB\,211211A, where the duration of the main burst complex is significantly longer than expected for neutron star merger systems, the system may be more `exotic' than the non-unity mass ratio model of \cite{gottlieb2023}, and may well be an indicator of a spiral armed accretion complex seen in simulations of 
neutron star mergers where only one of the stars carries a significant spin \citep{rosswog2024}, or neutron star magnetar mergers \citep{gao2022}. 
While such a scenario may seem at first sight far-fetched, it is worth remembering that the majority of observed neutron stars in globular clusters are millisecond pulsars. 
Based on observed systems, the study of \cite{rosswog2024} suggests that a non-negligible fraction ($\sim 5\%$) of merging systems can contain a millisecond pulsar.

Such mergers are predicted to produce a dominantly red kilonova with a weak or missing blue component; for the kilonova in GRB\,211211A, a significant blue component was observed -- although, it was not easily fit by kilonova models \citep{rastinejad2022}.
\cite{hamidani2024} noted that the blue and the red kilonova components are inconsistent with each other and they propose that the blue component is due to thermal emission from shock heated ejecta due to late engine activity.
In this scenario, the kilonova in GRB\,211211A is almost entirely red.

\subsection{The QPO at energies >25 keV only}
The data are split into two energy bands: low, and high (see Table\,\ref{tab:LCs}).
The relatively narrow low range is motivated by spectral fits that include a thermal component, with a cut-off power law plus blackbody having a thermal peak at 7.7\,keV and 13.3\,keV for a pure blackbody \citep{xiao2024}.
The first second of GRB emission was shown to have a low-energy break at $21.0\pm8.8$\,keV, when fit with two smoothly broken power-laws, and a blackbody temperature of $14.6\pm2.4$\,keV when a thermal component is included \citep{gompertz2023}.
In both cases, however, the thermal component is disfavoured when compared to a non-thermal model.

The `high' and `low' band periodograms in Figure\,\ref{fig:LSHS_low}, as well as the spectrograms in Figures\,\ref{fig:FERMI} \& \ref{fig:BAT}, show that the QPO at $\sim 22.5$\,Hz is dominant at energies $\gtrsim25$\,keV.
Despite the better fits for a purely non-thermal model \citep[e.g.,][]{xiao2024, gompertz2023}, given that the QPO is at energies $>25$\,keV, we suggest that the lack of QPO signal in the $15-25$\,keV band is due to the contribution of photospheric emission for the precursor.

\section{Conclusions}\label{s:conc}
We show that the candidate QPO is present in both the \fermi-GBM and the \swift-BAT data at energies $>25$\,keV via LS and STFT methods (see also PSD in Appendix \ref{app:A}).
By fitting a periodic light curve/pulse model to the precursor emission, we demonstrate that a periodic engine cycle is consistent with the QPO from the periodogram analysis -- and that periodic GRB pulses are preferred over a simple sinusoidal oscillation.
We further demonstrate, via an internal shock model based on shell collisions within the jet, the required minimum Lorentz factors for individual shells to produce the variability timescale of the precursor.
This analysis reveals a periodic engine duty cycle, responsible for the QPO, where the energy within each engine cycle is damped in comparison to the previous.
For the precursor, internal shocks occur only within the shells ejected within each engine cycle -- this preserves the periodicity of the engine within the emission and requires a rapid damping of the oscillation mechanism.

The minimum Lorentz factor from the pulse and peak-and-trough analysis requires values on the order of $\Gamma\sim100$; such a high Lorentz factor is inconsistent with a shock-breakout scenario for the precursor emission.
Via pulse peak-to-trough analysis, we find a minimum variability for the precursor shorter than the sampling timescale, $\lesssim 4$\,ms, and a mean variability timescale similar to that found by \citet{xiao2024}, $\sim15$\,ms.
The short variability timescale for the precursor is consistent with an origin within a relativistic outflow/jet -- an engine driven jet episode precursor to the more efficient, or powerful, main burst complex.
This suggests that the jet power within the precursor is coupled to the mechanism that produces the QPO.
The appearance of the QPO at $>25$\,keV in the precursor indicates that an additional emission mechanism exists at energies $<25$\,keV.
The contribution of emission from the jet photosphere at a temperature of $\sim$few keV would mask the non-thermal, pulsed lower energy emission within the jet.

The periodic signal between 18-25\,Hz contains $13.2\%$ of the precursor energy.
If the total energy is $E_{\rm pre, iso}\sim3.5\times10^{50}$\,erg (where we have assumed an emission efficiency of $\eta=0.02$, \S\ref{s:disc}), then the QPO has an energy budget of $E_{\rm QPO}\sim4.6\times 10^{49}$\,erg -- if the emission is isotropic, and $E_{j,\rm QPO}\sim7.5\times 10^{46}$\,erg for a bi-polar jetted outflow with a narrow $\theta_j\sim 3{\fdg}27 \sim 0.057$\,radian half-opening angle, where $E_{j,\rm QPO}\propto\theta_j^2$.
The observed luminosity of the precursor is $L \sim 2.5\times10^{49}$\,erg\,s$^{-1}$, and $L_j \sim 4.1\times10^{46}$\,erg\,s$^{-1}$;
for QPO coupled directly to the dipole power, then a QPO will have a luminosity of $L_{j,{\rm QPO}}\sim5.4\times 10^{45}$\,erg\,s$^{-1}$.
The opening angle used here is narrow when compared to the GRB population \citep[however, see][for a comparable narrow angle for short duration GRBs]{salafia2023}, where both long and short GRBs have a typical opening angle from afterglow jet-break modelling of $\theta_j \sim 5{\fdg}7 \sim 0.1$\,radian \citep{aksulu2022}. This is approximately a factor $\sim 7/4$ larger than the value inferred for GRB\,211211A \citep{rastinejad2022}, and so, if the precursor has a wider opening angle than the main emission, the energetics will be a factor $\sim 3$ larger.

We consider the implications under the assumptions that the central engine is a black hole or a magnetar:

{\bf For a black hole central engine:}
\begin{itemize}
    \item Lense-Thirring precession at the disk inner edge requires a radius of $5 - 9~r_g$ dependent on black-hole mass and spin.
    \item In the LT case, higher mass black holes $\gtrsim 4.5$\,M$_{\sun}$ with a spin $\chi\lesssim 0.7$ result in smaller radial requirements, $r\lesssim 7r_g$.
    \item The duration and damping timescale of the precursor is consistent with precession through a thin ring at the inner disk -- such a ring can indicate a warped and torn inner accretion disk region.
    \item Oscillation processes that work on Keplerian orbits cannot reproduce the observed QPO.
    \item MHD effects, such as MRI, can produce the observed oscillation. For inner disk oscillations, the disk density and the magnetic field are positively proportional, with disk densities of $\sim10^{12}$\,g cm$^{-3}$ requiring $B \sim 10^{14}$\,G.
    \item Such MHD-driven oscillations may also occur within the jet.
    \item For MHD oscillations, the damping timescale can trace the jet-disk coupling efficiency, or the crossing timescale for Alfv\'en waves set by the ratio of the disk height to wave velocity.
\end{itemize}

Where the jet and disk are coupled, and the QPO origin is LT of the inner disk, 
the smallest radial solutions $\lesssim6~r_g$ indicate a black hole mass of $M_\bullet \gtrsim 4$\,M$_{\sun}$.
A black hole mass in this range is unlikely to be the result of a binary neutron star merger, although such a system cannot be ruled out; we suggest that the analysis here points towards a black hole neutron star merger as the system progenitor \citep{davies2005}.

{\bf For a magnetar central engine:}
\begin{itemize}
    \item Where the QPO is directly coupled to magnetar spin period, the QPO will be at $\gtrsim$300\,km, and beyond the scale of a merger disk, $<200$\,km.
    \item Where the QPO is the result of MHD instabilities within the disk and set by a local Alfv\'en wavelength, then the QPO will be at $\sim20$\,km (or within the inner disk), for a disk density of $\sim 10^{12}$\,g cm$^{-3}$. The required magnetic field strength is $1 \leq B/ (10^{14} {\rm G}) \leq 3$ -- and consistent with a magnetar \citep{aguilera2025}.
    \item If the engine is a HMNS, the spin frequency is on the order of kHz, and therefore the QPO cannot be coupled to the HMNS rotation.
\end{itemize}
For long-lived and stable magnetars, the dipole radiation will inject significant energy into the kilonova -- such an engine-fed kilonova is not seen in the observations \citep{rastinejad2022, hamidani2024}.
Where the QPO is the result of Alfv\'en oscillations within the inner disk, then a short-lived ($\sim 0.25$\,s) magnetar or HMNS cannot be ruled out.
Neither can more exotic merger origins, such as a neutron star - pulsar within a globular cluster.

Although a magnetar or HMNS cannot be entirely ruled out, a black hole engine origin for the QPO is naturally explained via Lense-Thirring precession of the inner disk that is coupled to the jet at early times.
For an early turbulent accretion disk, the jet and disk will align via torques induced by Lense-Thirring, or the Blandford-Znajek process -- typically, the BZ process will dominate \citep{polko2017}.
The competing torques on the jet-disk system at early times can induce disk warping and tearing, where differential precession across a thin inner ring would reproduce the precursor timescale and characteristic damping.
The gap between the precursor and the main emission then reflects the timescale for the bulk disk to commence rapid accretion, powering the main GRB jet.
The gap timescale will be equivalent to the dynamical time for the bulk accretion at the disk inner-edge.

\section*{Acknowledgements}
We thank the anonymous referee for their insightful and helpful comments that have improved this work.
GPL thanks Jillian Rastinejad for useful comments on an early draft, Siong Heng for fruitful discussions, Mery Ravasio for a crash-course introduction to \fermi-GBM data, and Samuel Tosh and Michael Hebden for inspiring work on internal shock models during an undergraduate SURE summer internship at University of Leicester and a Royal Society funded summer undergraduate studentship at Liverpool John Moores University (respectively).

This work utilises tools from {\sc SciPy} \citep{2020SciPy-NMeth}, {\sc GDT-core} \citep{GDT-Core}, {\sc Nessai} \citep{nessai}, {\sc Bilby} \citep{bilby_paper}, {\sc Pride-palettes} \citep{pride}, {\sc HEASoft} \citep{FTOOLS}, {\sc Matplotlib} \citep{matplotlib} and {\sc Redback} \citep{sarin2024}.

GPL, CMBO and CT are supported by a Royal Society Dorothy Hodgkin Fellowship (grant Nos. DHF-R1-221175 and DHF-ERE-221005).
KLP acknowledges funding from the UK Space Agency. 
Dimple and BPG acknowledge support from STFC grant No. ST/Y002253/1. 
BPG acknowledges support from The Leverhulme Trust grant No. RPG-2024-117.   IM acknowledges support from the Australian Research Council (ARC) Centre of Excellence for Gravitational Wave Discovery (OzGrav), through project number CE230100016.
N. Sarin acknowledges support from the Knut and Alice Wallenberg Foundation through the "Gravity Meets Light" project and by and by the research environment grant ``Gravitational Radiation and Electromagnetic Astrophysical Transients'' (GREAT) funded by the Swedish Research Council (VR) under Dnr 2016-06012.
S. Rosswog has been supported by the Swedish Research Council (VR) under grant number 2020-05044 and by the research environment grant ``Gravitational Radiation and Electromagnetic Astrophysical Transients'' (GREAT) funded by the Swedish Research Council (VR) under Dnr 2016-06012, by the Knut and Alice Wallenberg Foundation under grant Dnr. KAW 2019.0112, by Deutsche Forschungsgemeinschaft (DFG, German Research Foundation) under Gemany's Excellence Strategy - EXC 2121 "Quantum Universe" - 390833306 and by the European Research Council (ERC) Advanced Grant INSPIRATION under the European Union's  Horizon 2020 research and innovation programme (Grant agreement No. 101053985).
RS acknowledges support from Leverhulme Trust grant RPG-2023-240.

\section*{Data Availability}
Data utilised are from public repositories, \href{https://www.swift.ac.uk/swift_portal/}{\swift data archive}, and \href{https://fermi.gsfc.nasa.gov/ssc/data/access/gbm/}{\fermi-GBM data products}.


\bibliographystyle{mnras}
\bibliography{ms} 



\appendix
\section{QPO signal double check}\label{app:A}
\fermi-GBM has multiple detectors -- for this work we utilise the two channels, n2 and na, the two highest signal-to-noise data sets for this event.
If the QPO is only present in one of these data sets, then the significance of the signal would be cast into doubt (note that \citet{xiao2024} performed comprehensive significance tests for this signal in \fermi, \swift, and {\it Konus-Wind} data and confirmed the signal).
However, we caution that the signal at 20-25\,Hz is seen within independent telescopes -- both \swift and \fermi here, and {\it Konus-Wind} in \citet{xiao2024} initial discovery.
The search in multiple detectors acts as a sanity check.

\subsection{Light-curve and Lomb-Scargle periodogram}
\begin{figure*}
    \centering
    \includegraphics[width=\textwidth]{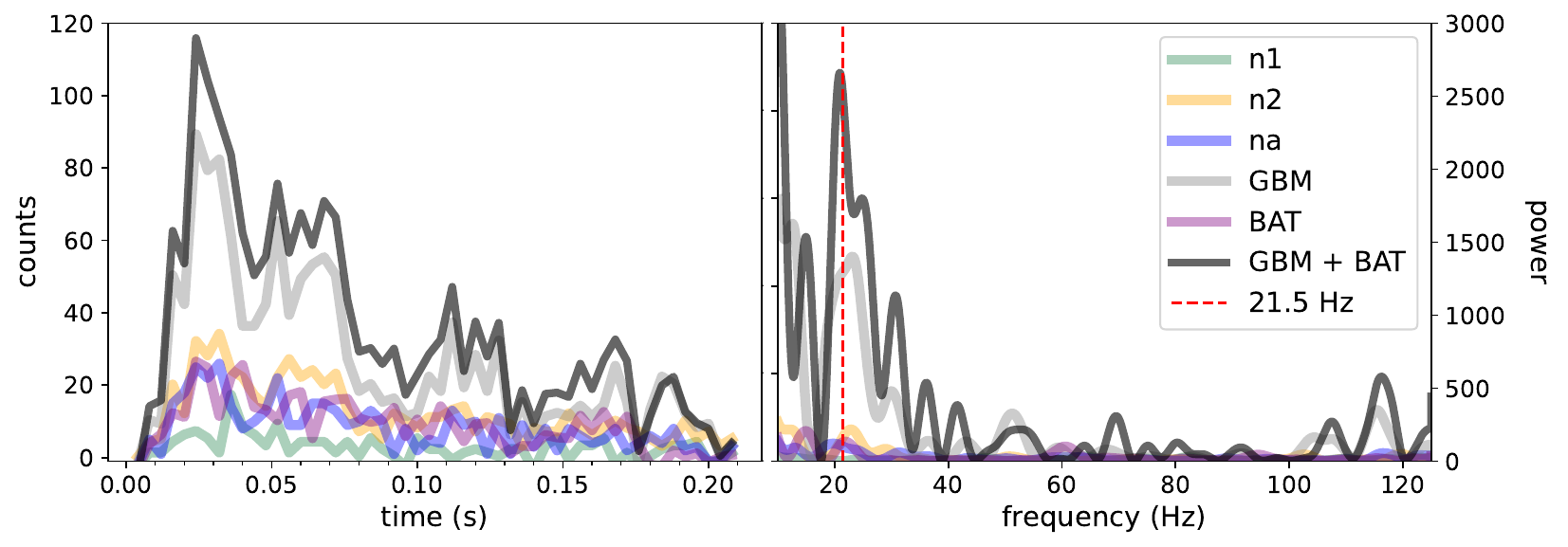}
    \caption{{\bf Left:} Light curves for n1, n2, na, all 12 GBM detectors summed, BAT, and GBM + BAT. {\bf Right:} Lomb-Scargle periodogram for each of the data sets shown in the left panel. A significant signal is seen at $\sim20-25$\,Hz. The red dashed line marks the 21.7 Hz frequency identified in the main body. The spectrum is cut at the low end at $<10$\,Hz, which is dominated by the precursor duration.}
    \label{fig:APP_A1}
\end{figure*}

Here we present the precursor light curves for 0.25\,s of data from \fermi-GBM n1, n2, na and all 12 detectors merged, plus the \swift-BAT data, and the sum of the merged GBM plus BAT data; the combined bands data is the `all' band (see Table\,\ref{tab:LCs} for band information), Figure\,\ref{fig:APP_A1}.
The detector light curves and LS traces for n0, n3-- n9, and nb are not shown, but are included in the 'all' GBM.

The left panel of Figure \ref{fig:APP_A1} shows the light curve data of the individual detectors and their sum, plus the other 9 detectors from \fermi-GBM.
The general FRED structure is maintained, and a clear `on/off' periodicity is evident on a $\sim20$\,ms timescale.
The right panel shows the periodogram of this signal.
The spectral power peaks at low-frequencies, which is tempting to attribute to red noise -- however, here the peak is that of the time series precursor duration, $\sim 0.25$\,s and $\sim 4$\,Hz.
Utilising a longer time series of data before and after (where after is more limited due to the main burst episode), reveals that the low frequency power spectra levels off at frequencies below $\sim 2$\,Hz, demonstrating behaviour inconsistent with red noise \citep{vaughan2005} -- long time series, lower frequency LS is not shown here.

\subsection{Welch's method}
\begin{figure}
    \centering
    \includegraphics[width=\columnwidth]{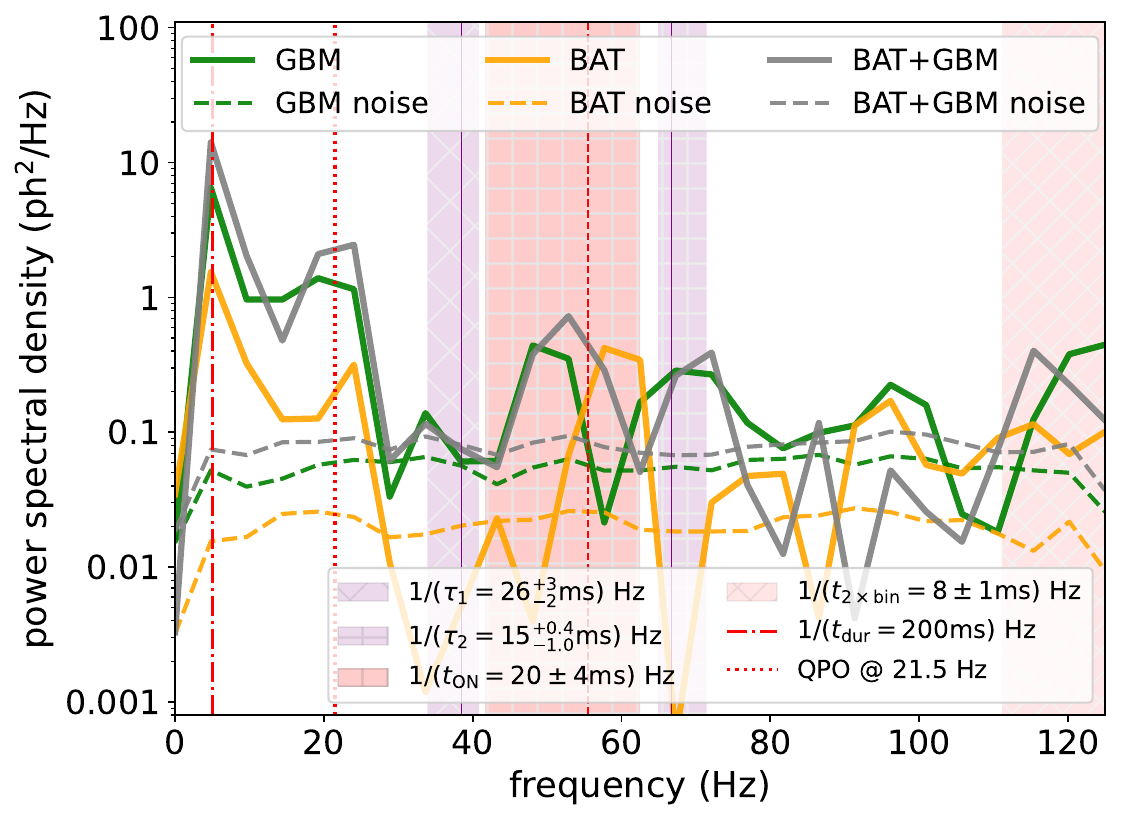}
    \caption{The PSD via Welch's method of the \swift-BAT, the \fermi-GBM na and n1, and the sum of the data. 
    The noise trace is that of the background data for a duration of 10\,s preceding the precursor. 
    Vertical lines and shaded panels indicate various timescales within the data: at $\sim5$\,Hz the red dash-dotted line corresponds to the precursor duration, $t_{\rm dur} = 0.2$\,s; 
    the 21.5 Hz line is the candidate QPO; 
    the shaded grid-hatched region at $\sim 55$\,Hz is the approximate `ON' cycle duration for the engine; 
    and the red cross-hatched region corresponds to the twice the sampling frequency. 
    The two purple shaded regions are the model (iii) pulse rise and decline timescales, with the shaded region at 38 Hz given by $\tau_1 = 26^{+3}_{-2}$\,ms, and at 67 Hz the $\tau_2 = 15^{+0}_{-1}$\,ms. 
    Note also that the MVT found by \citet{xiao2024} was 15 ms $\equiv \tau_2$.
    }
    \label{fig:welch}
\end{figure}

We further generate a power spectral density (PSD) using Welch's method \citep{welch1967} to confirm the results of the LS periodogram shown in the right panel of Figure\,\ref{fig:APP_A1}.
The PSD is shown in Figure\,\ref{fig:welch}.
The locations of the expected periodicity are given: the pulse duration of $\sim0.2$\,s; the proposed QPO at 21.5 Hz; the apparent `on' and `off' cycle time of half the QPO period; and the sampling frequency minimum variability timescale artifact are shown via red vertical lines and/or shaded regions.
In the PSD plot, the signal above background noise is seen via broad spikes at frequencies that are generally coincident with the cases noted.

The various spectral features are consistent with an inspection of the time series data.
A more detailed study of the precursor data is required to confirm the QPO significance \citep[e.g.,][]{hubner2022}, which is beyond the scope of the current work.

\section{Posterior distributions}\label{app:B}

\begin{figure}
    \centering
    \includegraphics[width=\columnwidth]{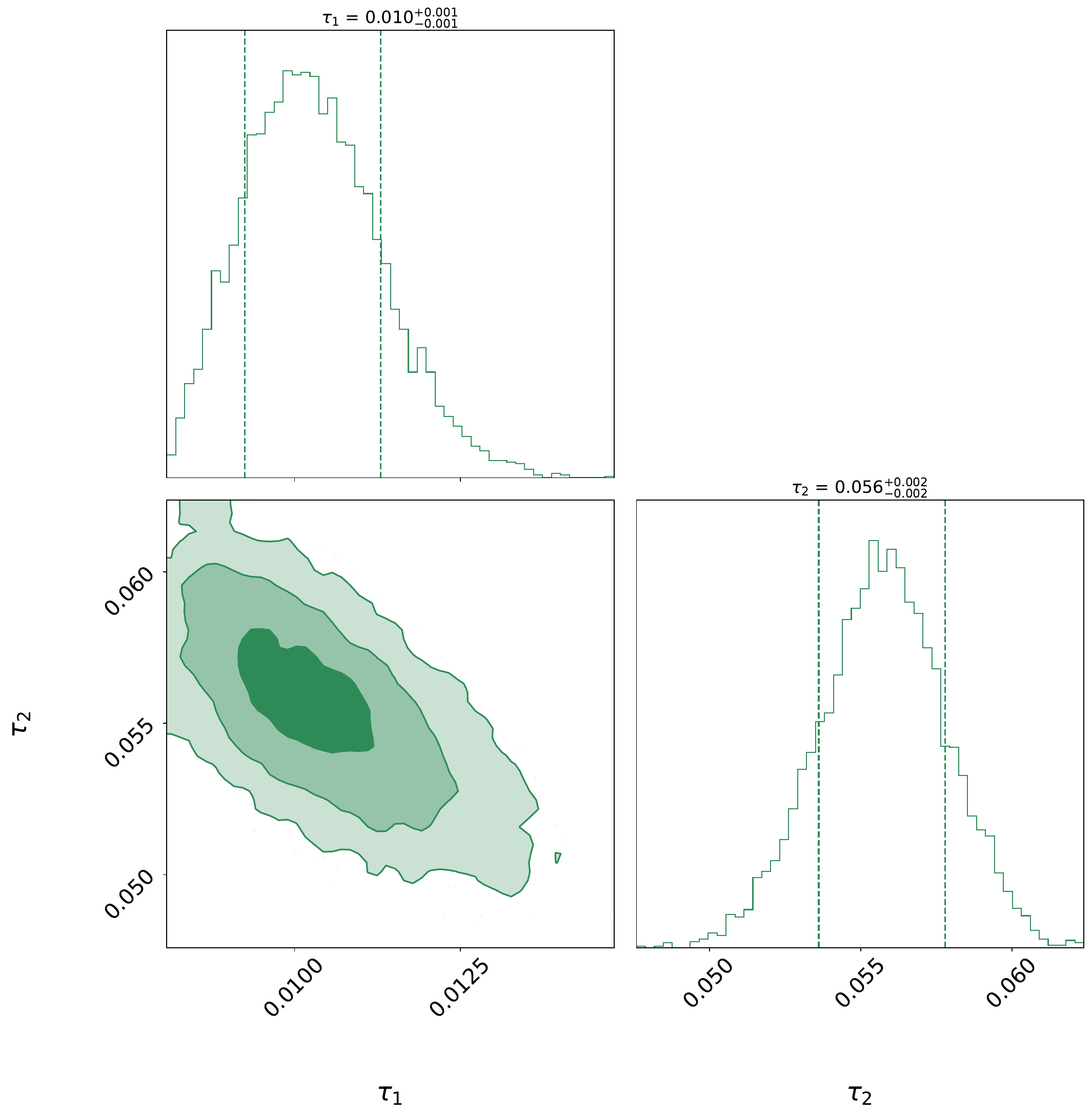}
    \caption{Model (i) -- single pulse.}
    \label{fig:single}
\end{figure}

\begin{figure}
    \centering
    \includegraphics[width=\columnwidth]{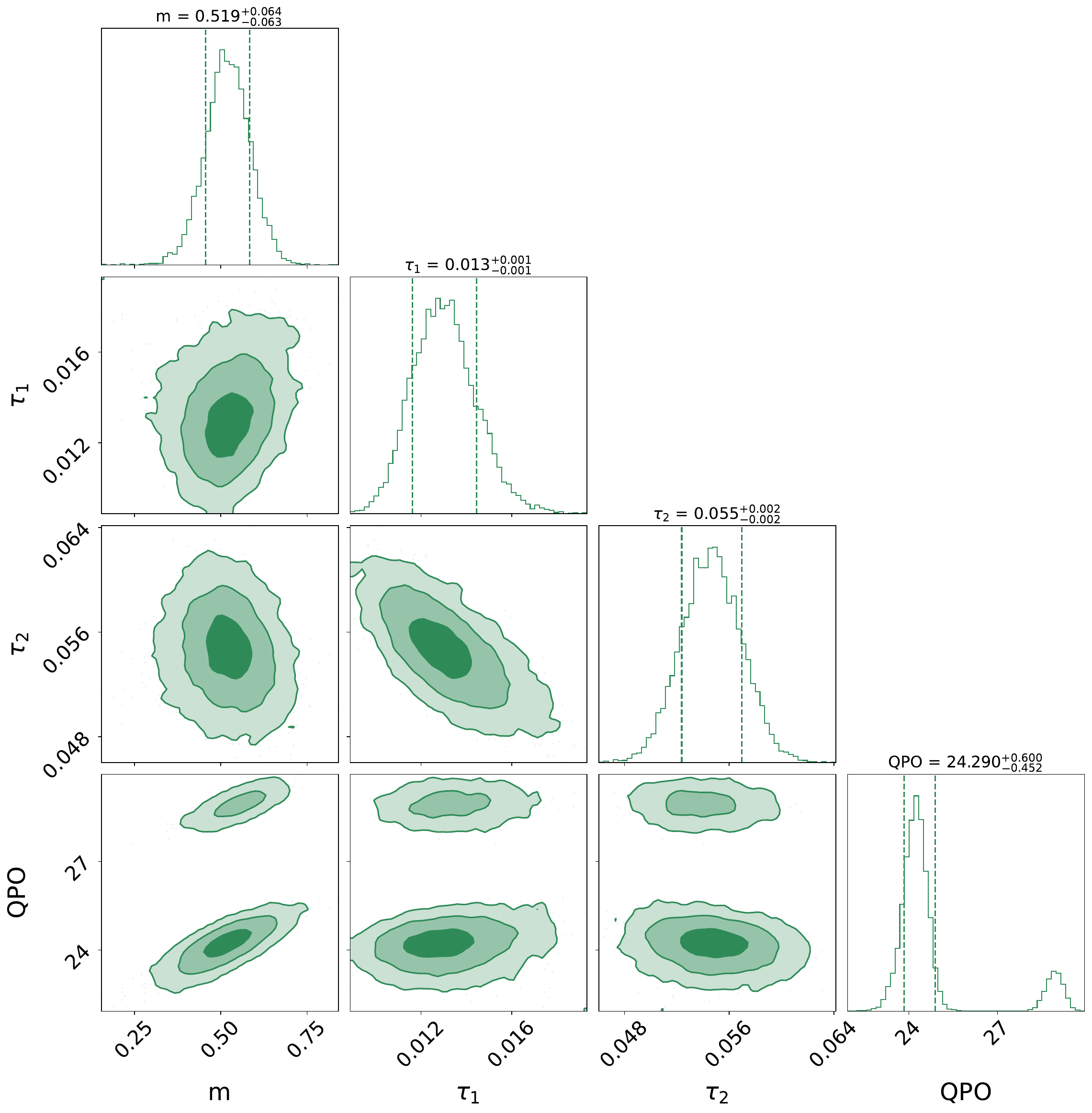}
    \caption{Model (ii) -- single pulse plus sinusoid.}
    \label{fig:singleQPO}
\end{figure}

\begin{figure}
    \centering
    \includegraphics[width=\columnwidth]{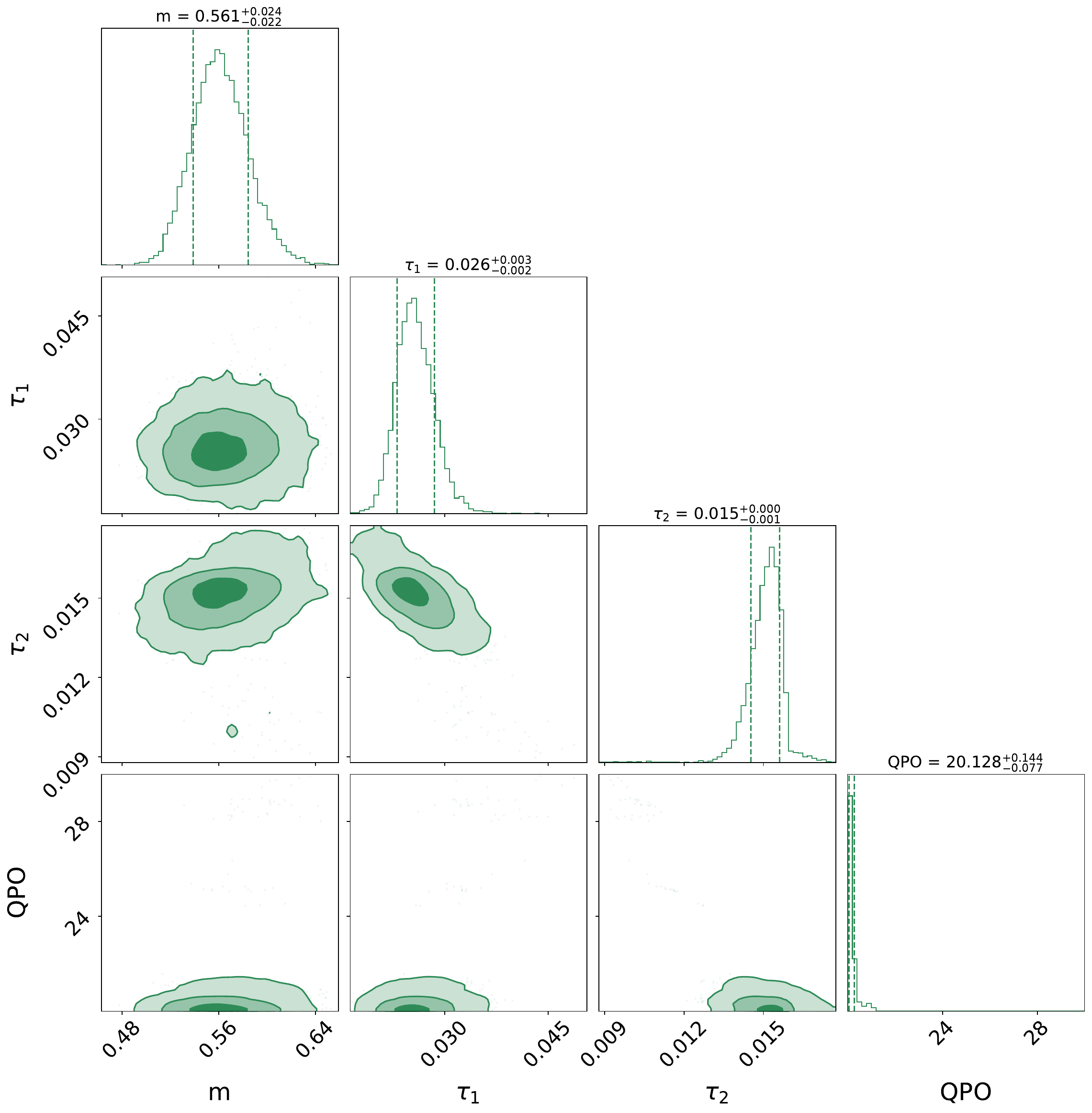}
    \caption{Model (iii) -- multiple pulses with periodicity.}
    \label{fig:multi}
\end{figure}

The model (i -- iii) fits to the `high' band combined \fermi-GBM and \swift-BAT precursor data, over a period of $0.25$\,s are shown graphically in Figures\,\ref{fig:single}, \ref{fig:singleQPO}, \& \ref{fig:multi}.

\bsp	
\label{lastpage}
\end{document}